\documentclass[12pt]{article}

\usepackage{latexsym,amsmath,amssymb,theorem,epsfig}
\usepackage{hyperref}
\DeclareFontFamily{OT1}{rsfs10}{}
\DeclareFontShape{OT1}{rsfs10}{m}{n}{ <-> rsfs10 }{}
\DeclareMathAlphabet{\mathscript}{OT1}{rsfs10}{m}{n}
\numberwithin{equation}{section}

\newcommand{\tr}{\text{tr}}

\def\a{\alpha}
\def\b{\beta}

\def\z{\psi}

\def\l{\lambda}

\def\p{\phi}

\def\r{\rho}
\def\s{\sigma}

\def\z{\zeta}

\def\D{\Delta}

\def\G{\Gamma}

\def\bb{{\bar b}}

\def\be{\begin{equation}}
\def\ee{\end{equation}}
\def\bea{\begin{eqnarray}}
\def\eea{\end{eqnarray}}

\def \td {\tilde}
\def \rx {{\rm x}}

\def \ha {{1 \ov 2}}
\def \sql {{\sqrt{\l}}\ }
\def \del{\partial}
\def \a {\alpha}
\def \aa {{\a'}}
\def\ov{\over}
\def \ci {\cite}

\def \foot {\footnote}
\def \bi{\bibitem}
\def\la{\label}\def\foot{\footnote}\newcommand{\rf}[1]{(\ref{#1})}

\def \no {\nonumber}
\def \adss {${\rm AdS}_5 \times S^5\ $}
\def \a {\alpha } 
\def \eps {\epsilon}

\def \N  {{\cal N}}

\def \iffa {\iffalse}
\def \tr  {{\rm tr\,}}
 
  \def \edd {\end{document}}
\def \N {{\cal N}}
\def \te {\textstyle}
\def \ha  {{\te {1\ov 2}} }
\def \be {\bea}
\def \ee {\eea} 
\def \aa  {{\rm a}}

\def \be {\bea}
\def \ee {\eea}
\def \cc {{\rm c}} 
 
\def \lm {v}

\def \tG {\Gamma} 

 \def \N  {{\cal N}}  \def \RR   {R^{(2)}}

\def \sql {\sqrt \lambda}
\def \str {{\rm s}}  \def \W  {{\cal W}}

\def \adn {{\rm AdS}_2}
\def \adt  {${\rm AdS}_2\ $}
\def \tot {{\rm tot}}
\def \p {\phi}
\def \xx {{\rm x}}
\def \rR   {{\rm R}}
\def  \WW   {{\rm W}}
\def \R  {R^{(2)}}
\def \ym  {{_{\rm  SYM}}}
\def \rG  {{\rm G}}\def \D  {{\rm D}}

\def \ed {\end{document}}
\begin{document}

\ \hfill{\small  Imperial-TP-AT-2020-05 }

\vspace{0.5cm}

\vspace{1.5cm}

\begin{center}

{\Large\bf  
Strong coupling expansion of 
circular Wilson loops \\
   \vspace{0.2cm}
   and string theories in AdS$_5 \times S^5$   and  AdS$_4 \times CP^3$
 }
 
\vspace{1.5cm}
{
Simone Giombi$^{a,}$\footnote{\  sgiombi@princeton.edu}\
 and
Arkady A. Tseytlin$^{b,}$\footnote{\ Also at the Institute of Theoretical and Mathematical Physics, MSU and Lebedev Institute, Moscow.
\\\hspace*{15pt} \ tseytlin@imperial.ac.uk}
}
\vspace{0.8cm}

{
\em \vspace{0.15cm}
$^{a}$Department of Physics, Princeton University, Princeton, NJ 08544, U.S.A. \\
\vspace{0.25 cm}
$^{b}$Blackett Laboratory, Imperial College, London SW7 2AZ, U.K.
}
\end{center}
\vspace{0.5cm}
\begin{abstract}
\noindent
We revisit the problem of matching the strong coupling expansion of the 
$ 1 \ov 2$ BPS circular Wilson loops in ${\cal N}=4$ SYM and ABJM gauge theories with their string theory duals in ${\rm AdS}_5 \times S^5$ and ${\rm AdS}_4 \times CP^3$, at the first subleading  (one-loop) order of the expansion around the minimal  surface. We observe that, including the overall factor  $1/g_{\rm s}$ of the inverse string coupling constant, as appropriate for the open string partition function with disk topology, and a universal prefactor proportional to the square root of the string tension $T$, both the SYM and ABJM results precisely match the string theory prediction. We provide  an explanation of the origin of the $\sqrt T$  prefactor based on special features of the combination of one-loop determinants appearing in the string partition function. The latter also implies a natural generalization $Z_\chi  \sim  (\sqrt T/g_{\rm s})^\chi$ to higher genus contributions with the Euler number $\chi$, which is consistent with the structure of the $1/N$ corrections found on the gauge theory side.
\end{abstract}

\newpage
\tableofcontents

\setcounter{footnote}{0}
\setcounter{section}{0}

\def \OO  {{\cal O}} \def \bG  {\bar \G}
\def \na {\nabla}\def \rr {{\rm r}}\def \cc  {{\rm c}}
\section{Introduction }
\def\theequation{1.\arabic{equation}}
\setcounter{equation}{0}

There is  a  history of attempts  to  match   gauge theory \ci{Erickson:2000af,Drukker:2000rr,Pestun:2007rz} 
and string theory  \ci{Berenstein:1998ij,Drukker:1999zq,Drukker:2000ep} results
 for the leading 
 terms in the  strong coupling  expansion of the expectation value   of 
 the  $\ha$ BPS circular Wilson loop (WL)   in $\N=4$ SYM theory   
  (see \ci{Kruczenski:2008zk,Kristjansen:2012nz,Buchbinder:2014nia,Forini:2015bgo,Faraggi:2016ekd,Forini:2017whz,Cagnazzo:2017sny}). 
The precise  matching was recently  achieved  for  the ratio
  of   the  $\ha$   and $1\ov 4$   BPS  WL expectation values  \ci{ Medina-Rincon:2018wjs}  (see also  \ci{Medina-Rincon:2019bcc,David:2019lhr}  for a discussion of similar matching in  the ABJM  theory \ci{Aharony:2008ug}). 
  However,  the direct    computation of the   string theory  counterpart  of  the expectation value of the  individual 
   WL, that 
  non-trivially depends on the normalization of the path integral measure, still remains a challenge. 
  
In the $SU(N)$   $\N=4$ SYM  theory    the Maldacena-Wilson operator  defined in the fundamental representation is  
given by 
$\W=  \tr  P e^{\int ( i A+\Phi)} $
 (note that we do not include  the usual $1/N$ factor in the definition of $\W$).
 Then   for a circular loop  one finds at large $N$ with fixed 't Hooft coupling $\l$  \ci{Erickson:2000af,Drukker:2000rr}: \ 
  $  \langle \W\rangle   =    N   { 2  \ov  \sql}  {I}_1 (  \sql)$. 
  Expanding at strong coupling, $\langle \W\rangle =  N  \l^{-3/4}  \sqrt{2 \ov \pi} \, e^{\sql} +...$. This 
  result should be reproduced by the \adss string perturbation theory   with  the string tension $T= {\sql \ov 2 \pi}= { \rR^2 \ov 2 \pi \a'}$, where here and below $\rR$ denotes the AdS radius. 
It was suggested  in \ci{Drukker:2000rr} that the pre-factor $ \l^{-3/4}\sim T^{-3/2} $   may have its origin  in the normalization of three    ghost 0-modes on the  disk (or the Mobius volume). 

This proposal, however, is problematic for several reasons. First, the effective   tension $T$   has its  natural origin  in 
  the string action 
  but   should not   appear in the   diffeomorphism volume or 
  the volume of residual Mobius symmetry.
Furthermore, the $ T^{-3/2} $  factor  (which would be universal  if related to the Mobius volume) 
would fail to explain the result for the $\ha$  BPS circular WL \cite{Drukker:2009hy} in the $U(N)_k \times U(N)_{-k}$ ABJM theory,
where the tension is  $T= \ha \sqrt{2 \l}$  (with $\l = {N \ov k}$) 
  while  the gauge theory  (localization)  prediction \ci{Kapustin:2009kz,Marino:2009jd,Drukker:2010nc} for the $\ha$ BPS Wilson loop
  in fundamental representation 
   is 
$  \langle \W\rangle   =  N  ( 4 \pi \l)^{-1} e^{\pi \sqrt{2 \l}}+...$. Note that, as above, in our definition we do not divide the Wilson loop operator by the dimension of the representation.\foot{\la{fo1} 
Our normalization of $\W$ in the $\ha$ BPS case corresponds in the localization calculation of \ci{Marino:2009jd,Drukker:2010nc} to computing the matrix model expectation value $\langle{\rm Str} {\tiny \begin{pmatrix}e^{i\mu_i}&0\\0&e^{-i\nu_j}\end{pmatrix}} \rangle$. Note that \cite{Drukker:2010nc} defines the Wilson loop expectation value by including an extra overall factor of $g_{_{\rm CS}}\equiv \frac{2\pi i}{k}$. Denoting by $ \langle \WW\rangle_{_{\rm loc} }$ the expectation value given in \cite{Drukker:2010nc}, we find that the strong coupling limit of the $\ha$ BPS Wilson loop in the ABJM theory is 
$ \langle \W\rangle = { 1 \ov g_{_{\rm CS}}}  \langle \WW\rangle_{_{\rm loc} }={ 1 \ov g_{_{\rm CS}}}
  \,  {1 \ov 2}   e^{\pi  \sqrt{2 \l}  + i \pi B}  
 =   { k \ov 4 \pi}  e^{ \pi \sqrt{2 \l} } =   { N \ov 4 \pi \l } \,  e^{\pi  \sqrt{2 \l} }  $, 
where we fixed the phase as  $B= \ha$.}

Another indication that the explanation  of the  prefactor     should be  different is that,  in  general,  one 
 expects  that the string counterpart of  the large $N$ term in 
$\langle  \W \rangle $ should be  the open-string partition function   on the disk, which should contain  an overall  factor of the inverse power of   the string coupling  (corresponding  to 
the Euler number $\chi=1$), i.e.
\begin{equation}\la{1111}
 \langle \W\rangle= Z_{\rm str}
 = {1\ov g_{\str} }
   {\rm Z}_1  +  \OO ( g_\str)  
   \, , \qquad\qquad     { \rm Z}_1 
   =  \int [dx]...  \ e^{ -  T \int d^2 \s\,   L }\,,
\end{equation}
where  $1\ov g_\str$   
 provides  the required overall factor of $N$.
 The fact that it is natural to define the WL  expectation value without  the   usual $1/N$ factor, and to include the 
 $1/g_\str$ factor in its string theory counterpart, was also emphasized in \cite{Lewkowycz:2013laa}.
 
In the $\N=4$  SYM  case we have \ci{Erickson:2000af,Drukker:2000rr}
\be \la{22}
g_{\str}  = \frac{g^2_{\rm YM}}{4\pi}=\frac{\l}{4\pi N}  \ , \qquad 
     \l = g^2_{\rm YM} N \ , \qquad   T
     = {\sql \ov 2 \pi} \ , \qquad  \langle \W\rangle  ={ N \ov  \l^{3/4}} \,  \sqrt{\tfrac{2}{ \pi}}\ e^{\sql}+... \ ,  
\ee
while in the ABJM case  \cite{Aharony:2008ug, Drukker:2010nc}\foot{
Here the ${\rm AdS}_4$ radius is 
$\rR= ( 2 \pi^2 \l)^{1/4} \sqrt{ \a'}$   with  $T\equiv  { \rR^2 \ov 2 \pi \a'}$. 
The shift $\l\to \l- {1 \ov 24}(1 - {1\ov k^2})$ in the  definition   of $\rR$ and thus of string tension 
\ci{Bergman:2009zh,Drukker:2010nc,Gromov:2014eha}
  is irrelevant to the   leading order in $1/T$   expansion that we   shall  consider below
 (as discussed in \ci{McLoughlin:2008he}, at  the 
  leading order we do not expect  renormalization of the relation for the string tension).
}
\be \la{55}
g_{\str}  = {   \sqrt \pi\,(2 \l)^{5/4}\ov  N}  \ , \qquad 
     \l = {N \ov k}  \ , \qquad   T = {\sqrt{2 \l} \ov 2}  \ , \qquad \ \ \ \  \langle \W\rangle  ={ N  \ov  4 \pi \l}\,  e^{\pi \sqrt{2 \l}} +...\ . 
\ee
Our central observation  is that both expressions for $\langle \W\rangle$   in \rf{22}   and \rf{55}     can be 
  universally  represented  as  
\be \la{100}
 \langle \W\rangle =  W_1 \,  \Big[1 + \OO (T^{-1})\Big] +   \OO ( g_\str)  \ , \qquad \qquad W_1 =  { 1\ov g_{\str}  }\,   \sqrt{\frac{ T}{ 2 \pi}}\  e^{   - \bG_{1}} \  e^{ 2\pi T  }  \ , \ee
 where $\bG_1$ is a  numerical constant.   
Below  we will argue that   \rf{100}  should be  the 
 expression for the  leading semiclassical  result for the disk string  path integral 
for a  minimal  surface in  ${\rm AdS}_3$     ending on a circle at the boundary (thus having  induced ${\rm AdS}_2$  geometry) 
  in the  ${\rm AdS}_n\times M^{10-n} $ string theory  with tension $T$ and coupling $ g_\str$. 
In \rf{100} the exponent $e^{ 2\pi T  }=e^{-I_{\rm cl}}$ comes from the value of the classical string action $I_{\rm cl}=  V_{\rm {\rm AdS}_2} \,  T  = -2\pi T$. 
 The constant $\bG_1$ comes  from  the  ratio of one-loop determinants 
 of string fluctuations near the minimal surface,  and  is  found to be 
      (see \ci{Drukker:2000ep,Kruczenski:2008zk,Buchbinder:2014nia}   and   section \ref{s2} below)
\be \la{01}
{\rm AdS}_5 \times S^5: \ \ \   \bG_{1} = \ha  \log (2 \pi) \ , \qquad \qquad \qquad \qquad 
{\rm AdS}_4 \times   CP^3: \ \ \   \bG_{1} = 0 \ . \ee
Including also   the $n=3$ case of 
${\rm AdS}_3\times  S^3 \times T^4  $  string theory,  one finds   for ${\rm AdS}_n\times M^{10-n} $   with $n=3,4,5$   that
   $\bG_{1} = \ha (n-4)   \log (2 \pi)$ (see  \rf{1000x}  below), and so  in general 
$W_1$ in \rf{100}  is 
\be 
\la{105} 
W_1 =  { 1 \ov ( \sqrt{ 2 \pi} )^{n-3} } \  { \sqrt T \ov g_{\str}  }\  e^{ 2\pi T  }  \ . \ee
Using \rf{01}  one can check that  the expression in \rf{100} or \rf{105}   is in   remarkable agreement with the 
gauge-theory expressions in \rf{22} and \rf{55}. 

As we explain below, it will also follow from our  argument  that  at higher genera (disk with $p$ handles with Euler number $\chi=1-2p$) 
   the $ \sqrt{T}$ factor in \rf{100}   should be replaced by $(\sqrt T)^\chi$, i.e. 
     the corresponding  term in the partition function should have a universal prefactor 
   \be   \la{1000}
 \langle \W\rangle =\sum_{\chi=1,-1, ...}  \cc_\chi  \,    \Big({{\sqrt T} \ov g_s}\Big)^\chi \,    e^{ 2\pi T  } \ \Big[1 + \OO (T^{-1})\Big] \ . \ee 
   This is indeed consistent with the structure of $1/N$ corrections found on the gauge theory side in \ci{Drukker:2000rr} and in \cite{Drukker:2010nc} (see section \ref{s4}).

It remains  to understand the origin of the simple prefactor $ \sqrt{\frac{ T}{ 2 \pi}}$  in \rf{100}.
In general,  the   expression for  such a   prefactor  in the  path integral  is  very sensitive  to the definition of path integral measure  which is subtle in string theory. 
In  section \ref{s3} below we  will  provide an  explanation for the  presence  of the 
$\sqrt T$   factor  starting from  the  superstring path integral in the static gauge 
\ci{Drukker:2000ep} 
(see also Appendix \ref{a1})
but we will not  be  able to determine  the origin of the  remaining ${1 \ov   \sqrt{2 \pi}}$ constant from first principles. 
This is  already a non-trivial result: since  the presence of this constant   is  fixed by the 
 comparison with the  SYM theory, we then have 
the string theory  explanation for  the  ABJM  expression in \rf{55}   (or vice-versa). 

In section \ref{sc3} we shall provide another consistency check of the universal expression for the  string partition function \rf{100} 
 by considering  the  analog of the familiar  soft dilaton insertion relation  and dilaton tadpole on the disk.

In  section \ref{s4}  we   will  emphasize the fact that 
 that  the universal  prefactor in  the disk partition function $\sim { \sqrt T \ov g_\str}$  in \rf{100} 
has a natural   generalization \rf{1000} 
 to higher orders   which is 
 consistent   with the structure of the $1/N$  corrections  found  on the gauge theory side. 
 We will  make some  concluding remarks 
 about some other WL examples  in section \ref{s6}.

It is interesting to note  that 
 the  factor  $ \sqrt{\frac{ T}{ 2 \pi}}$ in \rf{100}   looks  exactly  like  the one  associated with  just  one   bosonic   zero mode
(in  the standard  normalization of  the  path integral zero-mode measure, i.e. 
   $ {1\ov \sqrt{ 2 \pi \hbar}}$, $\ \hbar^{-1} = T$,  as was  used in a similar context    in  \cite{Medina-Rincon:2018wjs}).\foot{\la{fooo1} Here we   assume that the path integral measure   for a scalar field   is normalized so that  the gaussian integral  has a  fixed value  
$\int [dx]  \exp[ - { 1 \ov 2 \hbar} (x,  x) ] =1$, i.e.  $[dx]=  \prod_\s {dx(\s)\ov \sqrt{2 \pi \hbar}}$.
Then the  factor of string tension $T= \hbar^{-1}$  appears both in the measure and in the  action  and cancels out in 
the one-loop determinant expression  apart from possible  0-mode contribution.
}
In    Appendix  \ref{aa2}  we  will   discuss  a   possible origin of this zero-mode  factor, 
assuming  one starts  with the  disk path integral in conformal gauge 
where there is an  extra factor  containing  the  ratio of the   ghost determinant 
  and the determinant of the  two  ``longitudinal"   string coordinates subject to  
   ``mixed"  Dirichlet/Neumann   boundary  conditions, and thus admitting     conformal Killing  zero modes  discussed in Appendix \ref{b1}.

\section{One-loop string correction  in static gauge \la{s2}}
\def\theequation{2.\arabic{equation}}
\setcounter{equation}{0}


Let us 
  consider a circular WL  surface with ${\rm AdS}_2$ induced geometry, which resides in an ${\rm AdS}_3$ subspace of ${\rm AdS}_n \times M^{10-n}$, specifically:    

(i)  $n=5$: \ \   ${\rm AdS}_5 \times S^5$;  \ \ \   (ii) $n=4$: \ \  ${\rm AdS}_4 \times   CP^3$; \ \ \ 
(iii) $n=3$: \ \  ${\rm AdS}_3 \times   S^3\times T^4$.
\\
The string is point-like in the internal compact directions, satisfying Dirichlet boundary conditions. In general, the  planar  WL 
expectation   value    is   given 
 by the  string path integral with a disk-like world sheet ending on a circle at the boundary of AdS  space, 
 $ \langle \W \rangle = e^{-\tG}$, \ $\tG=\tG_0 + \tG_1 +  \tG_2 + ... $. Here  $\tG_0= - 2 \pi T $  is the  classical string action
 (proportional to  the renormalized  ${\rm AdS}_2$  volume $V_{\rm {\rm AdS}_2}= - 2 \pi$)  and 
 ${\tG_1}= \OO(T^0)$  is   given by sum of logarithms of  fluctuation determinants  (in which  we include  
  possible  measure-related  normalization factors).   
 
We shall    discuss  the  computation  of  the one-loop correction $\tG_1\equiv \tG^{(n)}_1 $  in  the  above ${\rm AdS}_n \times M^{10-n}$  cases following the heat kernel method  applied  in  the ${\rm AdS}_5\times S^5 $ case in 
\ci{Drukker:2000ep} and \ci{Buchbinder:2014nia}.
In this $n=5$ case 
the general form of the  static-gauge  
 string one-loop   correction   is  \cite{Drukker:2000ep}   
\bea 
\tG^{(5)}_{1}&=&\ha  \log  \frac{[{\rm det} (-\nabla^2 +2  )]^2 \  {\rm det} (-\nabla^2 +    \RR + 4   ) \  [{\rm det} (-\nabla^2 )]^5}
{[{\rm det} (-\nabla^2 +\frac{1}{4}\RR  +1) ]^8} \la{11}
 \\
&=&   \ha  \log  \frac{[{\rm det} (-\nabla^2 +2  )]^3 \  [{\rm det} (-\nabla^2 )]^5}
{[{\rm det} (-\nabla^2 +\ha ) ]^8}    \,. 
\label{bb1}
\eea
Here  we assumed  that the AdS  radius $\rR$ is scaled out   
and absorbed into the string tension $T= { \rR^2 \ov 2 \pi \a'}$
  so   that all operators are defined in the induced ${\rm AdS}_2$  metric  with radius 1  and    curvature 
   $\RR=-2$. We will come back to the radius dependence in section \ref{s3} below. 
In \rf{11}  we isolated the contribution of one special  transverse 
  ${\rm AdS}_5$  mode that, in general, is different from the other two:
this is the ${\rm AdS}_3$ mode  transverse 
to the  minimal surface 
(the other two transverse modes are transverse to ${\rm AdS}_3$),   see   \ci{Forste:1999qn,Drukker:2000ep}. In the present   case  of the   minimal surface  being  ${\rm AdS}_2$  we have  $\RR=-2$ so that its mass is actually the same as of the other two transverse ${\rm AdS}_5$ modes. 

Similar expression \rf{11}  is found in the 
 conformal gauge \cite{Drukker:2000ep},  
 provided   the contribution of  the  two   ``longitudinal" modes  cancels  as in flat space \ci{Fradkin:1982ge} against 
 that of  the  ghost determinant and Mobius volume factor 
   (modulo the 0-mode part of the longitudinal operator and related  definition of path integral measure, see  Appendix \ref{a1}
    for further discussion).

In the  less   supersymmetric  cases   with ${\rm AdS}_5 \to {\rm AdS}_n$  and $n=4,3$ 
 there are less massive bosonic  AdS   directions and
 part of the fermions  are  massless, i.e.  we   get   the following generalization  of \rf{11} 
\bea 
\tG^{(n)}_{1} &= & \ha  \log \frac{[{\rm det} (-\nabla^2 +2  )]^{n-3}  \    {\rm det} (-\nabla^2 + \RR + 4  ) \      [{\rm det} (-\nabla^2 )]^{10-n}}
{[{\rm det} (-\nabla^2 +\frac{1}{4}\RR +1) ]^{ 2n-2    }  \ [{\rm det} (-\nabla^2 +\frac{1}{4}\RR) ]^{10-2n} }
\label{b2}\\
&= & \ha  \log \frac{[{\rm det} (-\nabla^2 +2  )]^{n-2}  \ [{\rm det} (-\nabla^2 )]^{10-n}}
{[{\rm det} (-\nabla^2 +\frac{1}{2}) ]^{ 2n-2    }  \ [{\rm det} (-\nabla^2 -\frac{1}{2}) ]^{10-2n} }
\la{3}
\eea
 The fermion  masses   are controlled by  the   superstring    kinetic term  with  a projection matrix in the mass term. In the ${\rm AdS}_3 \times   S^3\times T^4$  case \ci{Drukker:2000ep} 
there are  4  massless  fermion  modes (which are partners of  $T^4$ bosonic modes)
  and 4 massive ones.
In the  ${\rm AdS}_4 \times   CP^3$  case one finds  \ci{Kim:2012nd}  that   there   are $2n-2=6$ massive  and 
$10-2n=2$ massless  fermionic  modes.



\

Let us first discuss the divergent part of  \rf{b2}    assuming  the standard heat-kernel 
regularization separately  for each determinant contribution. 
The  UV divergent part of  $\G_1= \ha \log \det ( -\nabla^2 + X) $  where  $-\nabla^2 + X$  is a scalar Laplacian 
 is given by ($\Lambda \to \infty$)
\be \la{1}
\G_{1,\infty}= - B_2 \log \Lambda \ ,  \qquad B_2 = \tfrac{1}{4 \pi} \int d^2 \s\,  \sqrt g\,  b_2 \ , \qquad b_2 = \tfrac{1}{6} \RR - X 
\ . \ee
Here we ignore boundary contributions (they contain a power of   IR cutoff   and are absent after renormalization of the ${\rm AdS}_2$ volume   or directly using the   finite value  for the Euler number of the  minimal surface).

In the case  of \rf{bb1}  we then find that in the total combination all $ \tfrac{1}{6} \RR$ terms cancel out 
(due to  balance  of    bosonic  and fermionic d.o.f.) 
and the   constant mass terms  also cancel  between bosons and fermions   so that 
 we are left only with contributions of $\RR$ terms from one special  bosonic  mode and  the 
fermionic modes
\be \la{2x} 
b^{(5)}_{2,\tot} = - \RR - 8  ( - \tfrac{1}{4}\RR ) = \RR \ ,   \qquad B^{(5)}_{2,\tot} =\tfrac{1}{4 \pi} \int d^2 \s\,  \sqrt g\,  \RR =\chi= \tfrac{1}{4 \pi} (- 2 \pi) (-2) =1  . \ee
For  general  $n$
the corresponding UV divergent part of \rf{b2} is  given  by the straightforward  generalization of \rf{2x}.
Again, all ${1\ov 6}  \RR$  terms in \rf{1} cancel  out  as do  the constant   mass terms   and   we find 
\bea 
&&b^{(n)}_{2,\tot} = - (n-3) 2  -( \RR  + 4) - ( 2 n-2) ( - \tfrac{1}{4}\RR - 1) - (10-2n) ( - \tfrac{1}{4}\RR)  = \RR \ , \la{2}  \\
 &&\qquad B^{(n)}_{2,\tot} = \tfrac{1}{4 \pi} \int d^2 \s\,  \sqrt g\,  \RR  =\chi=1 \ .\la{2211} \eea
The total result (coming  again just from  the $\RR$  terms in single bosonic mode and 8
fermionic  modes)  is  thus universal, i.e.  $n$-independent.  

Moreover,  the same result $B_{2\, \tot} = \chi$  for the coefficient of the UV divergence 
is found for  fluctuations near any minimal surface  (not even lying within ${\rm AdS}_3$) 
that has  disk topology (see \ci{Drukker:2000ep,Forini:2015mca}):
if $X$  is a ``mass matrix",   the contribution of 8  transverse bosons  is 
$b_{2b} = 8\cdot  {1  \ov 6} R^{(2)}-\tr X-R^{(2)}$ while of 8   fermions is 
$b_{2f} = 8\cdot {1  \ov 12} R^{(2)} + \tr X$  so that $b_{2\, \tot} = R^{(2)}$.

Note that in  general the Seeley   coefficient  is  
 $B_2= \zeta(0) + n_0 $    where $\zeta(0)$  is  the regularized number 
   of all non-zero modes  and  $n_0= n_b - \ha n_f $ is the effective  number of  all   0-modes (assuming fermions are counted as  Majorana or Weyl). 
   In the   present  static gauge  case   there are no obvious normalizable  0-modes
    (cf. remark below  \rf{bb10}), but we observe that the result  \rf{2211} is formally the same as what would come just from one ``uncanceled"  bosonic mode. 
    
The  universality of \rf{2211}  strongly suggests that 
the  mechanism of cancellation of  this  total ``topological" UV divergence should   also be universal. 
One  may     absorb it into the definition of the    superstring path integral measure  or cancel it against other 
   measure factors 
as discussed in the conformal gauge  in \ci{Drukker:2000ep}.\foot{\la{f1}
To recall,  the UV divergences   do not cancel automatically even in the bosonic  string theory in  flat space. 
The  combination of  $D$  scalar Laplacians 
  and the conformal ghost  operator $\Delta_{\rm gh}=P^\dagger P$  gives   (with all modes counted)   \cite{Fradkin:1981dd}
$B_2 = {1 \ov 4 \pi} \int d^2 \s\,  \sqrt g ({D\ov 6}  \RR - (  {2\ov 6} \RR + \RR) ={1 \ov 6} (D-8) \chi$. 
Assuming,   following  \ci{Alvarez:1982zi}, that  there are extra powers of  the UV cutoff in the  Mobius volume one divides over 
and in the  integrals over moduli, the net result is that one should add to the above $B_2$ an extra $\delta_{\rm top}   B_2 =  - 3 \chi = {\rm dim\, ker\, } P^\dagger -   {\rm dim\, ker\, } P$, thus getting $B_2 = {1 \ov 6} (D-8-18)   \chi= {1 \ov 6} (D-26)   \chi$. 
A similar argument  applies to  the   NSR string   where   $B_2 =  {1 \ov 4} (D-10)   \chi$. 
In the present  $D=10$   GS  superstring case there 
 is    an extra   conformal  anomaly/divergence from the  Jacobian of rotation from GS fermions to 2d fermions
(see \ci{gs-2d-div});
this  effectively amounts  to adding  3 extra massless fermion contributions 
 for each 2d fermion   contribution 
(or, equivalently, 
multiplying  the ${1\ov 6} \RR - {1\ov 4} \RR$ part of each fermion contribution to $b_2$ by 4); this gives 
$\delta_1  B_2 =- 3 \times 8 \times {1 \ov 4 \pi} \int d^2 \s\,  \sqrt g ({1\ov 6} \RR - {1\ov 4} \RR) = 2\chi $. 
\ 
 In the conformal gauge  the divergences from  the determinant  of  the 
ghost operator $(\Delta_{{\rm gh}})_{ab}= - g_{ab} \nabla^2 - R_{ab}$ 
cancel  against those of the determinant  of operator $\Delta_{\rm long}$  for 2  longitudinal scalars.
As in the bosonic case, one should also add $\delta_{\rm top}   B_2=-3\chi $ as explained above.
Summing   these  contributions   with \rf{2211}  gives 
$B_{2\, \tot} =B^{(n)}_{2, \rm tot} + \delta_1 B_2 + \delta_{\rm top}  B_2=  \chi  + 2\chi  - 3\chi =0\ .$  
}
An   
alternative is to use a  special  ``2d supersymmetric"  definition of  the one-loop 
 path integral  in the static gauge
(see below): 
   the cancellation of UV divergences is, in fact, automatic 
if one uses  a  
 ``spectral"  representation  for the total $\G_1$  
rather than     heat  kernel  cutoff for each individual determinant. 

\


Let us now turn to the   finite  part of  the one-loop effective action  in \rf{3}.
We will follow \ci{Buchbinder:2014nia}    which    completed  the  original computation  in  \ci{Drukker:2000ep} 
 of $\G_{\rm1}  $  in \rf{bb2}  
  based   on 
expressing   the determinants in \rf{bb1}  in terms of the  
well known  \ci{cam} heat kernels  of  the scalar and spinor Laplacians on  ${\rm AdS}_2$.
$\tG^{(n)}_1$   in \rf{b2}    contains  the  contributions   of the following ${\rm AdS}_2$  fields:
 (i)   $n-2$ scalars  with   $m^2=2$; (ii) 
$10-n$  scalars with $m^2 =0$; (iii) 
 $2n-2$  Majorana fermions with $m^2=1$; (iv)  $10-2n$  Majorana fermions with $m^2=0$.  
We  will  temporarily   set the ${\rm AdS}_2$
radius to  1  and  discuss  the dependence on it  later. 
  Let us first use the heat-kernel  
  cutoff for each  individual   determinant in \rf{b2}, i.e.  
\be \la{ab1} 
\ha \log \det \Delta = - \ha   V_{{\rm {\rm AdS}_2}} \int_{\Lambda^{-2}}^\infty { dt \ov t} \  K(t) \ , \ \ \ \ \ \ \ \ \ \ \ 
\  V_{{\adn}} = - 2\pi \  . 
\ee
The   trace of  heat  kernel  $K(t) $  for a   real  scalar and a Majorana 2d fermion  may be written as%
\be 
&&\qquad \qquad  K(t) = \tfrac{1}{2 \pi} \int_0^{\infty} d \lm\ \mu (\lm) \ e^{- t (\lm^2 +M)}\,, 
\label{4.7}\\
&&
\mu_b (\lm) =\lm \tanh (\pi \lm)\,, \quad M= \te \frac{1}{4}+ m^2\ ; 
\qquad \quad 
\mu_f (\lm) =- \lm \coth (\pi \lm)\,, \quad M= m^2 \ . 
\label{4.8}
\eea
 Here in $\mu_f$  we already accounted  for the negative sign  of the fermion contribution, so the total $K$  is just the sum of the bosonic and  fermionic  terms.
The  associated  $\z$-function is 
\be 
\z (z) = - { 1 \ov \G(z) } \int_0^{\infty} d \lm\,  \mu (\lm) \int_0^{\infty} d t\ t^{z-1} \ e^{-t(\lm^2+M)}= -
 \int_0^{\infty} d \lm\ \frac{\mu (\lm)}{(\lm^2+M)^z}\,. 
\label{4.10}
\ee
%
%
For example, for \adt scalars  
$\zeta(0) = B_2 = - \ha b_2= {1\ov 6}  + 
 \ha m^2$. 
The total  value of $\zeta(0)$ is  found  to be  1, i.e. the same as in \rf{2211}.
In general, the one-loop correction is 
\be \la{1200}
\Gamma_1 =\sum \ha \log \det \Delta=  -   \zeta_{\rm tot}(0) \log  \Lambda  + \bG_1 \ , \qquad 
\bG_1 \equiv  -\ha  \zeta'_{\rm tot}(0) \ , \qquad \qquad 
\zeta_{\rm tot}(0)=1 \ . \ee
\iffa
 Here we restored the dependence on \adt radius $\rR$  that  was assumed  to
be   1  in \rf{ab1},\rf{4.7}  (see also \rf{002}):  
if the \adt metric has  a scale $\rR$   then    the 
argument of the exponent in \rf{4.7} (with  dimensionless $v$)
 should have  an extra factor of $a^{-2}$   (so that $t$  and $\Lambda^{-2}$ in \rf{ab1} 
 have dimension of length squared). 
\fi
For  the  derivative  of the  scalar   $\z$-function  one finds ($A$ is the Glaisher constant) 
\be 
&&\z^{\prime}_b (0, M)=   - {\te {1 \ov 12}} ( 1+ \ln 2 ) +  \ln A    -  \int_0^{M} dx\, \psi \big(\sqrt{x}+\ha \big) \,, 
\label{4.21}\\
&& \z^{\prime}_b \big(0, \te\frac{9}{4}\big) = - \frac{25}{12} + { 3 \ov 2}  \ln (2 \pi) - {2 \ln A}\, , \qquad \qquad 
 \z^{\prime}_b \big(0, \te\frac{1}{4}\big) = - \frac{1}{12} + {1 \ov 2}  \ln (2 \pi) - {2 \ln A}\, , \la{2444} \ee
while  for  the massive fermion 
\be 
 &&\z^{\prime}_f  (0, M)=  - {\te { 1 \ov 6} }  + 2 \ln A      + {\sqrt{M}} + \int_0^M d x\, \psi(\sqrt{x})\,,
\label{4.30}
\\  &&  \z^{\prime}_f \big(0, \te  1 \big) =   \te \frac{5}{6}  -  \ln (2 \pi) + {2 \ln A}\, , \qquad \qquad  \z^{\prime}_f \big(0, \te  0 \big) = - {\te { 1 \ov 6} }  + 2 \ln A    
\  . \la{4} \ee
 The total contribution  to the finite part   $\bG^{(n)}_{1} $  in \rf{1200}
    corresponding to \rf{b2}   then found to have  a  simple form 
\be 
\bG^{(n)}_{1}&=&   - \ha \Big[ (n-2) \z^{\prime}_b  (0, \te\frac{9}{4})  + (10-n)   \z^{\prime}_b (0, \te\frac{1}{4})  + (2n-2)  \z^{\prime}_f (0, \te  1 )  +  (10-2n) \z^{\prime}_f (0, \te  0 )
         \Big]   \no \\
         &=&     \ha  ( n-4) \ln (2 \pi)\,. 
\label{118}\\
%
\bG^{(5)}_{\rm 1} &=&   \ha \ln (2 \pi) \ , \qquad\qquad  \bG^{(4)}_{\rm 1} = 0 \ , \qquad \qquad 
\bG^{(3)}_{\rm 1} =  -  \ha \ln (2 \pi) \ .
   \la{bb2} \ee
In the ${\rm AdS}_5\times S^5$ case ($n=5$)   the   computation   of the corresponding determinants   was also    carried out  using different methods   in \ci{Kruczenski:2008zk,Kristjansen:2012nz} 
with the finite  part    of the resulting expression for $\bG^{(5)}_1$   being  as in \rf{01}, \rf{bb2}.  
Note that  the  finite part  \rf{118}  happens to    vanish  in  the ${\rm AdS}_4 \times CP^3$ case ($n=4$). 

\

\iffa  
\foot{In ${\rm AdS}_3 \times S^3 \times T^4$ 
 case the result  in \rf{bb2}  is   $\G_1 =   - \ha   \ln (2 \pi)$, 
i.e.  is  opposite in sign to ${\rm AdS}_5 \times S^5$  one. 
Here   we  may hope  to compare to  strong-coupling limit of  WL   expression to be found in 
the dual 2d CFT described   in \ci{Sax:2014mea}.
  May be direct summation of graphs or localization may help get the result...  but there are issues...
  }
  \fi 
  
  
  It is interesting to note that  there exists a special  definition of  $\G_1$ in \rf{b2} 
  that automatically gives a UV   finite one-loop result.  
  Instead of  computing   separately  each determinant 
     let  us  use \rf{ab1}    and    sum up the  corresponding 
  spectral integral  expressions  
   under  a  common  integral over $v$ in \rf{4.7}.
    Interchanging the  order of  $t$- and $v$- integrals    and first integrating over  $t$ 
    we see that this integral is finite, i.e. the  proper-time  cutoff  is not required. 
   Using \rf{4.7}--\rf{4.8} we then get for  \rf{3} 
     \be 
&& 
 \bG^{(n)}_{1} = 
   \ha  {\te { V_{{\rm {\rm AdS}_2}} \ov 2 \pi}} \int_0^{\infty} d \lm\   \lm  \Big(
 \te   \tanh (\pi \lm) \big[ (n-2) \ln ( \lm^2+ {9\ov 4} )  + (10-n)  \ln ( \lm^2+ {1\ov 4} )  \big]\no\\
  && \qquad \qquad \qquad  \qquad \qquad  \ \  - \coth (\pi \lm) \big[ (2n-2) \ln ( \lm^2+ 1 )  + (10-2n)  \ln ( \lm^2 )  \big]\Big) \ , \la{001}
  \ee
  where  $ {{ V_{{\rm {\rm AdS}_2}} \ov 2 \pi}} =-1$.
  Remarkably,   the   integral over $v$   here 
    is  convergent at  both $v=0$    and $v= \infty$ (i.e. in the UV). In general, given the structure of the eigenvalues in (\ref{4.7})-(\ref{4.8}), one can see that convergence of the representation (\ref{001}) in the UV requires the sum rule $\sum_b (m_b^2+\frac{1}{4})-\sum_f m_f^2 = 0$, which is satisfied for the spectra in our problem. Evaluation of (\ref{001}) gives then a  finite   result   equal  to the one  in \rf{118}, i.e. 
  \be \la{1000x}
    \bG^{(n)}_{1}  =  \ha  ( n-4) \ln (2 \pi)\,.  \ee
  This prescription  of  not using proper-time cutoff   for individual $\log \det$ terms, i.e.  
    first combining the integrands  and then   doing the  spectral integral, may be viewed as a kind of  ``2d  supersymmetric" regularization.
    Indeed, the  balance  of the  bosonic   and fermionic degrees of freedom  in \rf{3}   suggests hidden \adt   supersymmetry
    \ci{Drukker:2000ep}.\foot{One implication  is  
      the vanishing of 
     the corresponding vacuum energy in \adt   observed in \ci{Drukker:2000ep}
    in the case of the strip parametrization $ds^2= {1\ov \cos^2\rho} ( dt^2 + d \rho^2)$, $\rho\in ( - {\pi\ov 2}, {\pi\ov 2})$.
    To recall, the  contributions of  a  scalar   with mass  $m^2_b$ and  a  fermion with mass $m^2_f$   to the ${\rm AdS}_2$  vacuum energy 
      are  \ci{Drukker:2000ep}\   $E_b (m^2) =- {1\ov 4} (m^2 + {1\ov 6})$   and  $E_f(m^2) = {1\ov 4} (m^2 - {1\ov 12})$
    so that for the spectrum in \rf{b2}  we get 
    $E_{\rm tot}= (n - 2) E_b(2)  + (10 - n) E_b(0)  + (2 n - 2)  E_f (1)  + (10 - 2 n) E_f(0)=  (n - 2) ( -{1\ov 2} - {1\ov 24}) + (10 - n) ( - {1\ov 24}) + (2 n - 2) ( 
   {11\ov 48}) + (10 - 2 n) ( - {1\ov 48})  =0$.
    }
    Then  the prescription of   combining the  spectral integrands of the  determinants together  
     may be   viewed as a  result of a  ``superfield"   computation  manifestly preserving 2d supersymmetry. 
Note however that, even though the integral in (\ref{001}) is finite, a dependence of $\Gamma_1$ on a normalization scale reappears on dimensional grounds if one restores the dependence on the radius $\rR$ inside the logarithms, as explained in the next section. This leads to an explanation of the $T$-dependent prefactor in (\ref{100}) and (\ref{1000}).

  \iffa 
Let us note that if we  re-introduce a   mass scale   $\mu= a^{-1}$   (related to the AdS radius)
 under the log's in \rf{001}  we  will find that  $\tG^{(n)}_{1} $ 
 depends on it via the same $\zeta(0)=1$  term as in \rf{1200}
 \be 
&&  \tG^{(n)}_{1,a } = 
   \ha {\te  { V_{{\rm {\rm AdS}_2}} \ov 2 \pi}}\, 8 \log (\rR^{-2}) \,  \int_0^{\infty} d \lm\   \lm  \, \big[
 \te     \tanh (\pi \lm)  - \coth (\pi \lm) \big] =-  \ha \log (\rR^2)  \ . \la{002}
  \ee
This illustrates that  as long as $\zeta(0) \not=0$ 
 the   finite part of the effective action is in principle scheme-dependent. 
 In general, the  scheme  is to be fixed by the condition of preservation of 
underlying symmetries. 
It  is natural to define $\Gamma_1$   in  this  special scheme where it is UV finite. 
\fi 


\iffa 
in < I >  we  have   Box G     or  delta(z,z)  or    tr < z | exp - Box | z > 
So this corresponds to   just summing up K(t=eps)   in 2.21   in current 19.6  file with relevant coeffs
or replacing  log ( v^2  + ...)   by   exp[ - eps ( v^2 + ...) ].
\fi

\def \bbG {{\bar G}}  \def \bx  {\bar x} \def \rx {{\rm x}}

\section{Dependence 
on  AdS radius:  origin  of the $\sqrt T$ prefactor \la{s3}
 }
\def\theequation{3.\arabic{equation}}
\setcounter{equation}{0}

Let us now   explain the presence of the 
$\sqrt T = { \rR \ov \sqrt { 2 \pi \a'}}  $ prefactor in the string one-loop partition function \rf{100}.
As the  definition   of  quantum string path   integral  (in particular, integration measure)  is 
subtle and potentially    ambiguous  our aim is to identify the one that is consistent 
with underlying symmetries  and AdS/CFT duality. 

In the previous   section we  ignored the dependence of the one-loop correction on the AdS  radius $\rR$. 
Let  us now discuss   how the string path integral may depend on  it. 
Let us start with the  classical string action in ${\rm AdS}_n$ of radius $\rR$. One  possible 
 approach is to rescale the  2d fields so  that 
the factor  of $\rR^2$   appears in front of  the  action
\foot{For example, 
starting with $ds^2 = dr^2 + e^{2r/\rR} dx_i dx_i   $ we get 
$ ds^2 = \rR^2 ( d\bar r^2  + e^{2 \bar r}  d \bx_i  d \bx_i)$.
  Note that   after the rescaling  the tension $T$ and  coordinates $\bx^m$ are dimensionless.}
\be 
 I &=&\tfrac{1}{2}  T_0  \int d^2 \s\,\sqrt g\,   G_{mn} (x)\, \del^a x^m \del_a x^n + ... \la{003}\\
&=& \tfrac{1}{2}  T  \int d^2 \s\, \sqrt g\,    \bbG_{mn} (\bx)\, \del^a \bx^m \del_a \bx^n + ...    , \qquad 
T= \rR^2  T_0, \quad \ T_0=  \tfrac{1 }{ 2 \pi \a'} \ . \la{007}
 \ee
 Using    either \rf{003} or \rf{007}   the  expression  for one-loop correction will depend also on the assumption 
 about the path integral measure. 
 If the measure is defined covariantly the final result  should  be the same.

 Let  us   consider  the path integral   defined  by \rf{003} 
 in terms  of the original unrescaled coordinates $x^m$ 
of  natural dimension of length, so that  $G_{mn} (x) $ is dimensionless
and depends on the  AdS  scale   $\rR$. 
The  string  $\s$-model  path integral may be defined  symbolically  as (cf. footnote \ref{fooo1})
\be\la{771} 
Z= \int \prod_{\s,m}  \, \sqrt{\tfrac{ T_0}{ 2\pi}}\,   \sqrt { G(x(\s))}\,  [dx^m(\s)]\ldots \ \exp \big[ - \ha T_0 \int d^2 \s \,\sqrt g \,   G_{mn} (x)  \, \del^a x^m \del_a x^n + ...\big] \ . \ee
Expanding near the minimal surface ending on the boundary circle we will get the  induced  \adt  metric depending on the 
same  curvature scale   $\rR$ as $ G_{mn}$. 
 Then rotating the fluctuation fields   to the 
tangent-space components $\tilde \rx^r$   and also rescaling them by  $\sqrt{ T_0}$    (so that they will   be normalized 
as  $|\tilde \rx |^2 = \int d^2 \s \sqrt g \, \tilde\rx^r \tilde\rx^r$)
we will find that   the 1-loop  contribution 
from  a single scalar  is 
$Z_1=  (\det \Delta )^{-1/2}$   where $\Delta= - \nabla^2 + ...$  depends on the induced \adt metric
and  has  canonical dimension of  (length)$^{-2}$   with eigenvalues  scaling as $\rR^{-2}$. 
In the heat kernel  representation  
$\G_1=-\log Z_1 =  \ha \log \det \Delta = - \ha \int^\infty_{\Lambda^{-2}}   { dt \ov t}\,  \tr \exp ( - t \Delta) $ 
the parameter $t$  and the cutoff  $\Lambda^{-2}$   will now  have  dimension of (length)$^{2}$  and  we will get
   instead of  \rf{1200}   (cf.  \rf{1},\rf{2211})
\be 
\G_1 = - \zeta_{\rm tot}(0) \log (\rR  \Lambda) + \bG_1 
 \ , \qquad\qquad      \zeta_{\rm tot}(0)=\chi=1  \la{317}
\ee
As discussed  in  section  \ref{s2}, 
   the    UV  divergence   is expected to   be cancelled  by an extra  ``universal" contribution   $\log (\sqrt{ \a'} \Lambda)$ from the  superstring measure  
    (see  footnote \ref{f1}).  
    We assume that this   universal contribution
(depending only on the Euler number of  the world sheet but not on details of its metric) 
may only  involve the string scale  $\sqrt{ \a'} $
but not the  AdS radius.
As a result,   $\G_{1\, \rm fin} = -\chi  \log {\rR \ov  \sqrt{ \a'}}  + \bG_1$. The argument of  log  is thus  $\sim (\sqrt T)^\chi$, i.e. 
  we get 
\be \la{3077}
Z \sim e^{-\G_{1}}\ \  \to \ \   \big(\sqrt T \big)^{ \zeta_{\rm tot}(0)}=( \sqrt T )^\chi= \sqrt T  \ .
 \ee
This explains the origin of the $  \sqrt T$   factor in the disk partition function   \rf{100}.

As was noted   below \rf{2211},  the coefficient of the UV   divergent term in \rf{317} is, in fact,  
the same   for all minimal surfaces with disk topology   and thus  the dependence of the string partition 
function  on the   scale $\rR$   or effective tension $T$  through the $\sqrt T$     factor in \rf{3077} should be universal.  This means, in particular, that the  factors $1/g_{\rm s}$   and $\sqrt T$ in \rf{100}
will cancel in the ratio of expectation values of different Wilson loops with disk topology. 
Moreover, the fact that the power of $T$ in (\ref{3077}) is controlled by the Euler number $\chi$ implies that at higher genera, for a disk with $p$ handles, we should find  that $\langle \W\rangle$ includes the universal prefactor $(\sqrt{T}/g_{\str})^{\chi}$ as in (\ref{1000}). This is in precise agreement with the large $N$ expansion of the localization results both in ${\cal N}=4$ SYM and ABJM cases, as we explain in more detail in section \ref{s4}.

The   result of adding  the  above  universal counterterm  $\log (\sqrt{ \a'} \Lambda)$  is equivalent to  just 
defining   the  one-loop  partition function to be UV finite  by  first   combining  all the contributions  using 
 the spectral representation \rf{001}. There we set $\rR=1$   and  to  restore the
 dependence on the radius 
 $\rR$ of the \adt metric  we need  to  add   the    mass scale   factor  $\rR^{-2}$  
 under the logs in \rf{001}  (cf. \rf{ab1},\rf{4.7}). 
 To make  the argument of the logs dimensionless  we  also need to introduce 
 some normalization scale $\ell$ (i.e.   $\log \det \Delta \to \log \det ( \ell^2  \Delta) $
 or, equivalently, add $\ell$ factor in the path integral measure). 
 Then    we  find that  $\tG^{(n)}_{1} $  in \rf{001}   
 depends on  $\rR$ via the same $\zeta_{\rm tot}(0)=1$  term as in \rf{1200},\rf{317}, i.e.  via  an extra 
 contribution  (to be added to \rf{1000x})  
 \be 
  \delta \tG^{(n)}_{1 } = 
   \ha {\te  { V_{{\rm {\rm AdS}_2}} \ov 2 \pi}}\, 8 \log (\rR^{-2}\ell^2 ) \,  \int_0^{\infty} d \lm\   \lm  \, \big[
 \te     \tanh (\pi \lm)  - \coth (\pi \lm) \big] =-   \log (\rR\, { \ell^{-1} )}    \ . \la{002}
  \ee
The dependence on $\ell$  illustrates the fact 
 that  as long as $\zeta_{\rm tot}(0)\not=0$,  the one-loop  contribution, even if  defined  to be  UV finite 
  by the spectral   representation (or  some  analytic  regularization like the  $\zeta$-function one   \ci{Hawking:1976ja}), is  still  scheme (or measure) dependent.  
Choosing $\ell\sim  \sqrt{\a'}$,  which is here  an obvious   choice  in the absence of any other  available scales
(and  which is  also   suggested  by the $T_0$ dependence in \rf{771}), we  again end up with the  required result \rf{3077}.
  
We shall   discuss  some   other    approaches to  the derivation  of 
the dependence of the  one-loop correction on $T$ in  the next section and  Appendices  \ref{a1}  and \rf{aa2}.

\section{
Dilaton insertion and  derivative over  gauge  coupling    \la{sc3} }
\setcounter{equation}{0}
\def\theequation{4.\arabic{equation}}

As  another check of  consistency   and universality of the expression \rf{100} 
  for the 1-loop string partition function  for a minimal surface  
   with disk topology, let us  consider a closely related   object --  the  insertion 
   of the    dilaton   operator in the expectation value  or the  dilaton  tadpole on the disk  with  WL  boundary conditions.
   Here   we  shall explicitly consider the  SYM case but a similar discussion should apply also to the ABJM case.

Let us first  recall   the zero-momentum dilaton insertion relation, or the familiar ``soft dilaton theorem"   in flat space. 
The dilaton  $\p$  couples  to the string   as \cite{Fradkin:1984pq} 
\begin{equation}
I=
 \int d^2  \s \sqrt g \big[  \ha  T_0  G_{mn} (x) \del^a  x^ m \del_a  x^n +   \tfrac{ 1 }{ 4 \pi} \R \p(x) \big]\ , 
\end{equation}
where $T_0 = { 1 \ov 2 \pi \a'}$. The string-frame  metric $G_{mn}$  expressed in  terms of the  Einstein-frame  metric  in $D$ dimensions 
 is  $G_{mn} = e^{{4\ov D-2} \p}\bar G_{mn}, \ \  \bar G_{mn} = \delta_{mn} + h_{mn}$
 and  thus the  (zero-momentum) dilaton vertex operator   in flat space is (cf.  \ci{Tseytlin:1986ti,deAlwis:1985zy})\foot{The canonically   normalized  dilaton field $\bar \p$ 
 that appears in the generating functional for scattering amplitudes, i.e.  having  the same  kinetic term  as the graviton 
 in the effective action,   $S\sim \int d^D x \sqrt{ \bar G} [   - 2 \bar R +  \ha  (\del \bar \p)^2  + ...]$,
 is related to $\p$ as $ \bar \p = {4 \over  \sqrt{D-2} }\p$  so that 
 $I= I_0  -  \bar V_0 \bar \p + ..., \ \  \bar V_0  = - {1\ov \sqrt { D-2} } (  I_0  + {D-2\ov 4} \chi )$.  
 Note also that in \rf{c1}  we ignored  possible boundary term  as  its   role usually is only to ensure the
 coupling to the correct value of the Euler number. }
 \be \la{c1}
 I= I_0  -  V_0 \p + ... \ , \ \ \ \ 
 V_0 =- \tfrac{4}{ D-2}  \int d^2 \s \sqrt g \Big [  \ha  T_0    \del^a  x^m \del_a  x_m  +  \tfrac{D-2}{ 4}  \tfrac{ 1 }{ 4 \pi}  \R   \Big] =- 
 \tfrac{4}{ D-2}  I_0  - \chi  \ , \ee 
where $ I_0 = \ha T_0   \int d^2 \s \sqrt g \,    \del^a  x^m \del_a  x_m$ and $   \chi= { 1\ov 4 \pi} \int d^2 \s  \sqrt g \R$. 
Since the expectation value of the action $I_0$    may be    obtained by applying  $ - T_0 { \del\ov \del T_0 } $ 
to the string path integral (cf. \rf{37}),  
the  insertion  of the  zero-momentum dilaton into the  generating functional  for scattering amplitudes  $ Z = \int [dx]  e^{- I_0 + V_0 \p +   V_h h + ...} $ 
is then given by (here $ \langle  1 \rangle $=1)
\be\la{c2}   
{\del \ov \del \p}  \log Z  =  \langle  V_0 \rangle   = - \tfrac{4}{ D-2}   \langle   I_0 \rangle   -  \chi  =
 \tfrac{4}{ D-2}     T_0 { \del\ov \del T_0 }\log Z     -  \chi   \ . 
\ee
In the standard  cases  of a  bosonic   closed   string  or open string with Neumann boundary conditions 
 there are $D$ 
 constant 0-modes, so one  finds   $ Z \sim T_0^{D/2} $    and  $ \langle   I_0 \rangle  = - \ha D $ (assuming ``covariant" 
 regularization  in which $\delta^{(2)}(\s,\s)=0$, see   \cite{Tseytlin:1988ne}).
 The same relation   is true  also  for   the  fermionic string     as the number of bosonic  translational 0-modes  remains the same. 
 
 In the  superstring case   ($D=10$)    for the  tree-level topology   of a disk  ($\chi=1$)
 eq. \rf{c2}  reads 
 \be\la{c3}   
{\del\ov \del \p} \log Z   =  \langle  V_0 \rangle   = - \ha   \langle   I_0 \rangle   -  \chi  =  \ha  T_0 { \del\ov \del T_0 } \ln Z   -1 
  \ . 
\ee
Adapting this relation  to  our present case of  fixed contour boundary conditions  with the   expectation value of the action given by 
 $  \langle   I \rangle  = - \ha    $  (see \rf{391})  the  analog of \rf{c3}   becomes 
 (including  in  $  \langle   I \rangle$  also the classical   contribution  of  an \adt minimal surface  $ \langle   I \rangle_{\rm cl} =  T ( - 2 \pi) = - \sql $)
  \be\la{c4}   
{\del \ov \del \p}\log Z   =  \langle  V_0 \rangle   = - \ha   \langle   I \rangle   -  1  =  \ha (  \sql   +   \tfrac{1}{2} -2) 
 =\ha     \sql  -  \tfrac{3}{4} 
  \ . 
\ee
 Since   the   constant part of the 
 dilaton is  related to the  string coupling  which itself  is related to  the SYM coupling as in \rf{22}, i.e. 
 $ g^2_{\rm YM}= 4\pi g_\str= 4 \pi  e^\p$,  we  may compare  \rf{c4}  to the derivative of the  circular WL  expectation value with respect to the coupling constant on the gauge theory side. 
 The  normalized gauge theory path integral is defined  by 
 $ \langle ...\rangle_\ym  \sim    \int [dA...]  e^{-  S_{\rm SYM}} ... $,  \ $ \langle 1 \rangle_\ym
  =1$  where
 \be \la{cc4}
 S_{\rm YM} =\int d^4 x \, L_\ym\ , \qquad \qquad 
 L_\ym       =  { 1 \ov 4 g^2_{\rm YM}} \,  \tr ( F^2_{mn} + ...) \ .   \ee
 We  assume that  
  the metric is Euclidean and the $SU(N)$ generators  are normalized  as  $\tr(T_i T_j) = \delta_{ij}$.
 Since   the factor in front of the action is 
 \be \la{000}
  e^{-\p } = g_\str^{-1} = { 4 \pi  \ov  g^2_{\rm YM}}=  { 4 \pi  N   \ov  \lambda } \ , 
  \ee
   the   derivative  over the  constant part of the dilaton $\p$ corresponds   on  the gauge-theory   side 
  to the  insertion of the  SYM  action into a  correlator.  In particular,  in the case of the WL expectation value 
 (here and below $ \langle \W \rangle \equiv  \langle \W \rangle_\ym $)
  \be \la{cy4}
 {\del \ov \del \p}  \log   \langle \W \rangle =  {   \langle S_\ym   \W \,    \rangle  \ov  \langle   \W \,    \rangle}=    \l  {\del \ov \del \l}  \log  \langle \W \rangle   \ . 
  \ee
 Since  the gauge-theory result  at strong coupling is 
  $\langle \W\rangle  =  N  \l^{-3/4}  \sqrt{2 \ov \pi} \, e^{\sql} +...$   we conclude that 
  \be \la{ccc4}
 {\del \ov \del \p}\log   \langle \W  \rangle =         \l  {\del \ov \del \l}  \log  \langle \W \rangle  = \ha  \sql  -  \tfrac{3}{4}   + ...  \ , 
  \ee 
  which   is  in agreement   with  the string   theory expression   \rf{c4}. 
  
  Note that while in the string theory relation  \rf{c4} we used  that the insertion of the string action is given by derivative over  the tension, 
   on the gauge theory  side  a  similar  relation \rf{cy4}  involves   differentiation over   the gauge coupling.  
   The two are in agreement because  on the string side the dependence  on $\l$   comes  from both the 
   dependence on the tension and  also dependence on the string coupling (the $-\chi=-1$ term in \rf{c3},\rf{c4}).
   Thus, once again,  one  needs  the  independent $1\ov g_\str$   and $\sqrt T$ factors  in the string theory   disk partition function 
   \rf{100}  in order  to have the consistency between the dilaton derivatives, or  equality of the dilaton insertions on the string and  gauge theory  sides. 
   
 The  above   discussion   has a natural generalization  to  the   string 
  partition function  on a  disk with handles  or  $1/N $ corrections on  the gauge theory side. 
  For a  surface of Euler number $\chi$,   using    \rf{399} we get    the  following analog of \rf{c2} 
  generalizing \rf{c4}
\be\la{c42}   
{\del \ov \del \p}  \log Z  =  \langle  V_0 \rangle   = - \ha    \langle   I \rangle   -  \chi  =
 \ha     T { \del\ov \del T }\log Z     -  \chi  =  \ha  \sql -   \tfrac{3}{4}    \chi     \ ,
\ee
where we used that $T_0  { \del\ov \del T_0 } =  T { \del\ov \del T }$. 
The subleading term  $-   \tfrac{3}{4} \chi$ 
   is consistent   with the general form  of the prefactor
    $Z_\chi \sim   ({\sqrt T \ov g_\str})^\chi \, e^{2 \pi T}$ in \rf{1000}.  
  Indeed,  note that $g_\str = e^\p$   and that  switching to the Einstein-frame metric (cf. \rf{c1}) 
   corresponds to 
   $ T \to  e^{{1\ov 2} \p} T$ (cf. \rf{c1}), so that ${\sqrt T \ov g_\str} \sim  e^{- {3\ov 4} \p} $. 
   This is in agreement  with the gauge-theory side  since  
   the dependence on the  dilaton  is directly correlated as in \rf{000},\rf{cy4} 
    with  the dependence on $\l$ (which appears only as a  factor in front of the 
   SYM action), while the  dependence on $N$   may come  not only 
   from the factor \rf{000}  in the action    but also  from traces  in higher order  gauge-theory  correlators.
   Indeed, according to  the gauge-theory result  (see \rf{444})  the  genus $p$ term in $\langle \W \rangle$ 
   depends on  $\l$  as  $ \l^{{6p-3\ov 4} }= \l^{- {3\ov 4} \chi}$.
   Equivalently, \rf{c42}  follows simply  from the fact that 
   $\l {\del \ov \del \l}\log Z = (\ha  T {\del \ov \del T}   +    g_\str {\del \ov \del g_\str } ) \log Z $. 
   
One can also perform a further   consistency check   by considering a
direct generalization of the above   relations to the case of the  local  (i.e. ``non zero-momentum") 
   dilaton  operator insertion. On the gauge theory side the derivative over a local coupling or  local 
   dilaton   is essentially the Lagrangian in \rf{cc4}
   and one finds  \ci{Danielsson:1998wt,Fiol:2012sg,Hovdebo:2005hm}\foot{   
   Eq. \rf{c5}   is a direct counterpart  of  the   exact   form of  the correlation function of the $\ha$ BPS Wilson loop with the $\Delta=2$ chiral primary operator   which is a special case of 
    the correlator of the Wilson loop and the $\Delta=J$ CPO
     first obtained in \cite{Semenoff:2001xp}.
   The function $f(\lambda)$ also appears in the so-called Bremsstrahlung function \cite{Correa:2012at}.
   The 
   dilaton operator ${\cal O}_{4}$  is  a descendant of the $\Delta=2$   chiral primary,  i.e. ${\cal O}_{4} \sim 
   \tr  (F^2 + \Phi D^2 \Phi + ...)$  and is different   from the canonical 
   form of the SYM   Lagrangian  $L_\ym$    in \rf{cc4}  by  a total derivative term  (in  conformal correlators one may further drop the terms  proportional to the  scalar and spinor    equations of motion as they  produce only contact terms     \ci{Klebanov:1997kc,Liu:1999kg}).
      Note   that   we use Euclidean notation (as, e.g.,  in \ci{Hovdebo:2005hm})   and  in our  normalization   \ci{Liu:1999kg}\ 
   $\langle   L_\ym(x)  L_\ym(x' )    \rangle = {3 N^2 \ov  \pi^4   (x-x')^8}$.  
   }
   \be \la{c5}
 {\delta \ov \delta \p(x)}\log  \langle  {  \W }  \rangle  = {  \langle  { L_\ym(x)} \,   \W \rangle\ov \langle \W \rangle}
 = - {1  \ov  8 \pi^2  \, d_{\perp}^4} f(\l)   \  ,\\
  f(\l) =    {\l }   {\del \ov \del  \l} \log  \langle  \W  \rangle =   \ha   \sql      { I_2(\sql) \ov I_1(\sql) }
 = \ha  \sql - \tfrac{3}{ 4}   + ... \ . \la{cc5}
 \ee
 In \rf{c5}   we assume   that  dependence on the local dilaton 
   is introduced   by  $  L_\ym \to  e^{-\p(x)} L_\ym$  and  $\phi$ is set to be constant as in \rf{000} 
   after the differentiation. 
 In \rf{cc5}   we used that $\langle  \W  \rangle = {2 \ov \sql} I_1 ( \sql)$, i.e. $ f(\l) $  is the same 
  function that appeared also in \rf{ccc4}.
  For a WL    defined   by a 
   circle of unit radius on the $(x_1,x_2)$-plane centered at the origin,
    the position dependent factor $d_{\perp}$ in (\ref{c5}) is given explicitly by (see, e.g., \ci{Berenstein:1998ij,Alday:2011pf,Giombi:2018hsx})
\begin{equation}\la{ddd}
d_{\perp}=\tfrac{1}{2}\sqrt{(r^2+h^2-1)^2+4h^2}\,,\qquad 
r^2=x_1^2+x_2^2\,,\qquad 
h^2=x_3^2+x_4^2\,.
\end{equation}
One can verify that integrating (\ref{c5}) over the position $x=(x_1,x_2,x_3, x_4)$ of the operator insertion, using the regularized expression for  the integral\footnote{To evaluate this integral, one may,  for instance, 
 first integrate over $r$, then integrate over $h$ and  finally remove the power divergence at  $h=\epsilon\to0 $, i.e. 
 $\int d^4x\, \frac{1}{d_{\perp}^4} = { 2 \pi^3 \ov \epsilon} - 8 \pi^2  + {\cal O}(\epsilon)  \to -   8 \pi^2 $.  } 
\begin{equation}
\int d^4x\ \frac{1}{d_{\perp}^4} =(2\pi)^2 \int^\infty_0 dr\, r  \int^\infty_0 dh\,  h\ \frac{16}{\big[(r^2+h^2-1)^2+4h^2\big]^2}=-8\pi^2\,,
\end{equation}
one recovers the   relation  (\ref{cy4}). 

On the string theory side, the  corresponding  local  dilaton operator is 
 (cf. \rf{c1}; here $D=10$  and $L _\str= \ha    z^{-2}  \del^a  x'^m \del_a  x'^m + ...$ is the \adss  superstring Lagrangian)
 \be 
 &&\la{c6}V(x)  =-    \int d^2 \s \sqrt g \big(  \tfrac{1}{2}  T  L_\str  +    \tfrac { 1 }{ 4 \pi}  \R   \big) \,  K(x-x'; z) \ , \quad \\
 &&\la{c666}
 {K(x-x'; z)} = \, c_4\,  {  z^4 \ov [z^2   + (x- x')^2]^4 } \ , \ \qquad   c_4 = {   \Gamma(\Delta)\ov \pi^{\frac{d}{2}}\Gamma(\Delta
 -\frac{d}{2})}\Big|_{_{d=4, \Delta=4}}  = \tfrac{6 }{ \pi^2} \ . \ee
  $K$ in \rf{c666}  is the    bulk-to-boundary   propagator of the massless  dilaton in ${\rm AdS}_5$  ($\Delta=4$). 
 Integrating over  the  4-dimensional boundary coordinates  gives  back     $V_0$   that appeared in   \rf{c4} 
 (indeed,  $\int d^4 x \,   K(x-x'; z)  =   2  \pi^2  { 1 \ov 12} c_4 =1  $). 
 Note that 
 the  correlator in \rf{c5} is to be compared to the 
string theory dilaton  insertion on the disc  with the dilaton vertex  operator  defined relative to the Einstein-frame metric 
so that  the 2-point functions of the graviton and dilaton (and the corresponding dual operators)   are decoupled.

Note that the normalized correlator (\ref{c5}) for the case of the WL  corresponding to  a    
 straight line  is related to the one for the circle by a conformal transformation, and it takes the same form as (\ref{c5}), with the same function $f(\lambda)$, and $d_{\perp}$ being simply the distance from the straight line (i.e., for a straight line along the $x_1$ direction, $d_{\perp}=\sqrt{x_2^2+x_3^2+x_4^2}$) \ci{Berenstein:1998ij}.  
Using 
  the \adt  surface  in the  straight line case  ($z= \s, \  x^0=\tau, \ x^i=0$)  we get for the contribution of the
  leading classical term and the 
   $\R=-2$ term in $V$ in \rf{c6}:
  \be 
\langle  V_{{\rm cl} +\R} (x) \rangle &=&-    \int d^2 \s \sqrt g \,  ( \ha   T   +     \tfrac { 1 }{ 4 \pi}   \R \big)   \,   K(x-x'; z) \no\\
 &=&-  c_4   \tfrac { 1 }{ 4 \pi}  (\sql - 2)  \int^\infty_{-\infty} 
  d\tau \int ^\infty_0 \,   { d \s  \ov \s^2}   {  \s^4 \ov (\s^2   + \tau^2 + d_{\perp}^2 )^4 }  =- \tfrac{1}{16  \pi^2\,  d_{\perp}^4} ( \sql -2) \ .  \la{c7}
 \ee
Then the string theory  expectation value  $ {\delta \ov \delta \p(x)} \log Z  =   \langle V(x)  \rangle$  indeed  matches \rf{c5} if one adds in the  last bracket in \rf{c7} an extra $+\ha$  coming from   the  1-loop  quantum fluctuations of the bosonic and fermionic string coordinates  in $\langle L_\str \rangle$, in parallel to what happened in   \rf{c4}.

 \iffa 
But this is essentially what I was suggesting with the d/dT logZ argument, and it should again involve the \delta(0) calculation that we discussed...But maybe it's useful to try to think of this more generally for expansion around an arbitrary solution with disk topology, rather than focusing on AdS2 propagators (and maybe not assuming static gauge throwing away longitudinal).
I think you were basically doing such calculation in that PLB paper, though it seems around eq. (20) you are actually setting \delta(0)=0 and keeping only the zero mode contribution? (In our case, if we do include the longitudinal coordinates in dxdx, then we'd have 3 zero modes from those giving a -3/2, and then again we would need to explain an extra +1 coming presumably from GS fermions...).
\fi 


\section{Universal form of   higher  genus  corrections  \la{s4}} 
\def\theequation{5.\arabic{equation}}
\setcounter{equation}{0}

An important   feature of the $\sqrt T \ov g_\str$   prefactor in \rf{100}  is that  it  has a  natural generalization 
$({\sqrt T \ov g_\str})^\chi$ 
to  contributions from higher genera 
\rf{1000} (cf.  also \rf{c42},\rf{399}). 
Let us  recall that  in the case of the  $SU(N)$  $\N=4 $  SYM 
theory  the exact expression for the expectation value of  the  $\ha$ BPS  circular WL $\W = \tr P e^{\int (i A + \Phi)}$  expanded at large $N$ and then at  large $\l$  is 
\ci{Drukker:2000rr,Pestun:2007rz}  ($ L^1_{N-1}$ is the  Laguerre  polynomial)\foot{
\la{fo3}   Let us note that this expression applies to the  SYM theory  with  the $U(N)$  gauge  group;  the result   in the $SU(N)$ case  
is obtained by  multiplying \rf{444}   by $\exp ( - {\l \ov 8 N^2} )$  \ci{Drukker:2000rr}. 
This   factor expressed in terms of $g_\str$ and $T$ in \rf{22}   is $\exp ( - { g^2_\str \ov 2 T^2})$
and thus   is subleading compared to $H$ in \rf{446} at large $T$; we therefore   ignore it here. 
}
\be \la{444}
 \langle \W \rangle = \ e^{ \lambda \ov 8 N}  \,  L^1_{N-1} ( - \tfrac{\lambda}{4 N}) 
 =  N \sum_{p=0}^\infty   \, \tfrac{  \sqrt 2   }{  96^p\,  \sqrt \pi \,  p!}\,     
   {\lambda^{{6p-3\ov 4} }\ov N^{2p}}\, e^{\sql} \, \Big [ 1 +   \OO(\tfrac{1}{ \sql}) \Big] \ . 
\ee
It was suggested in   \ci{Drukker:2000rr}  that the sum over $p$ 
  may be   interpreted as a  genus    expansion on the string side. 
  Remarkably, 
we  observe that once the overall  factor of $N$   is included, i.e. one considers the expectation value of 
$\tr ( ...) $  rather  that ${1\ov N} \tr ( ...)$,  
   the  full dependence on $N$ and $\l$  in the prefactor of $e^{\sql}$  in  \rf{444}
 combines  just  into    $ (N \l^{-3/4})^{1-2p}$.
Rewriting  \rf{444}  in terms of the  string tension $T= {\sql \ov 2 \pi}$   
and string coupling $g_\str = {\l \ov 4  \pi N}$  as defined in \rf{22}  we then get 
\be \la{445}
 \langle \W \rangle  
 =   \sum_{p=0}^\infty   \cc_p \,  \Big( { \sqrt T \ov g_\str}\Big)^{1-2p}    \, e^{2 \pi T } \, \Big [ 1 +   \OO({T^{-1}}) \Big] \ , \ \ \ \ \ \ \ \ 
\cc_p
=\tfrac{1 }{ 2 \pi\,    p!} \, (  \tfrac{ \pi }{12})^p \ , 
\ee
which is the same as \rf{1000}  where $\chi= 1-2p$ is the Euler number  of a disk with $p$ handles. 

Furthermore,   the  sum  that  represents  the  coefficient of  the leading large $T$  term in \rf{445} 
 has  a simple closed expression: since  
  $  \sum_{p=0}^\infty   \cc_p\,   z^p = {1 \ov 2 \pi} \exp ({\pi \ov 12} z ) $ we find 
   as in    \cite{Drukker:2000rr} (see  also \ci{Okuyama:2006ir})
\be \la{446}
 \langle \W \rangle  
 =   e^{ H }\, W_1\   \Big [ 1 +   \OO({T^{-1}}) \Big]    \ , \qquad \qquad 
 W_1= { \sqrt T \ov  2 \pi g_\str}  \, e^{2 \pi T } \ , \qquad   H\equiv  { \pi   \ov 12 }\,   { g^2_\str \ov T}\ . 
\ee
Here    $W_1 $ is  the   leading  large $N$ or   disk  contribution    in  the SYM theory given by 
\rf{100}  (with $e^{-\bar \G_1}= \sqrt{ 2 \pi}$    according to \rf{01}).
$H$   may be interpreted as representing a  handle insertion operator, i.e.
higher  order  string  loop corrections  here simply   exponentiate.   
Such  exponentiation 
  is expected in the ``dilute handle gas" approximation  of thin far-separated handles
which should  be relevant  to the leading order   in the  large tension
  expansion considered in \rf{444},\rf{445}  (cf. \ci{H}).

It has  another interesting   interpretation  suggested in \ci{Drukker:2005kx}. If one  considers  a circular Wilson loop in the totally symmetric rank $k$ representation of  $SU(N)$  then for  large  $k,\, N$  and $\l$ with $\kappa= { k \sql \ov 4 N}  =   {k\, g_s \ov 2\,  T}=$fixed  its  expectation value 
should  be  given by the exponent  $\exp ( - S_{\rm D3})$ of the  action of the classical D3-brane solution. 
In the limit  of $1 \ll k \ll  N $  this  description should apply also to the case of the WL  in the $k$-fundamental representation described by a   minimal  surface  ending on a
multiply wrapped circle and   here one  finds  \ci{Drukker:2005kx} that  $ S_{\rm D3} =N f(\kappa) = 
 - k \sql   - { k^3 \l^{3/2} \ov 96  N^2}  + \OO ({ k^5 \l^{5/2} \ov   N^4}) $.  If one {formally}   extrapolates this expression to 
 $k=1$, i.e. a single circle case  discussed above,   then      it becomes\foot{   
   This  corresponds  to a  resummation of the expansion in \rf{444}  for fixed $ {  \sql \ov N}$, i.e. 
      when $g_s \sim T $ are both large \ci{Drukker:2005kx,Okuyama:2006ir}   which is formally  a different limit compared to the one discussed above    and leading to \rf{446}.}
 $ S_{\rm D3} = - 2 \pi T  -  { \pi   \ov 12 }\,   { g^2_\str \ov T}   + \OO(   { g^4_\str \ov T^3}   )$, i.e.  $\exp ( - S_{\rm D3})$
   reproduces  precisely   the  exponential  factor  $e^{2 \pi T + H }$  in \rf{446}.

A similar  structure  \rf{445} of the topological expansion  should appear 
   in  the case of the $\ha$   BPS circular   WL  in the ABJM theory  which was   computed from localization in  \cite{Drukker:2010nc}.
    According to \rf{55},  in that case  we have 
${ \sqrt T \ov g_\str} = { N \ov \sqrt{ 8 \pi} \, \l} = {k\ov \sqrt{8 \pi} } $ 
 where $k$ is the CS level 
so that  $ \langle \W \rangle  $   should   be  a    series in 
$ ({ \sqrt T \ov g_\str})^\chi \sim k^\chi \sim  ( {1 \ov g_{_{\rm CS}}}  )^{\chi}$ (cf. footnote \ref{fo1}).
Translating the leading and  the first subleading $1/N$ corrections  to the WL expectation value  found explicitly   in  
 \cite{Drukker:2010nc} into our notation we get\foot{\la{fo2}
 Note that we use the notation $g_{_{\rm CS}}  \equiv { 2 \pi i \ov k}$   for what was called $g_s$ in 
 \ci{Drukker:2010nc} in order  not to confuse it   with the type IIA string  theory coupling $g_\str$ in \rf{55}.
The leading correction in eq. (8.19) in   \cite{Drukker:2010nc} is to be multiplied by  $g_{_{\rm CS}} ^2$
according to the definition of the topological expansion in (8.1) there. 
Also, as already mentioned in footnote \rf{fo1}, 
  with our  definition of the WL expectation value  $ \langle \W\rangle =\langle \tr ( ...) \rangle =
 { 1 \ov g_{_{\rm CS}}}  \langle \WW\rangle_{_{\rm loc} }$, 
where    $\langle \WW\rangle_{_{\rm loc} }$ is the   gauge theory localization  expression of   \cite{Drukker:2010nc}.}
\be \la{447}
\langle \W\rangle  =\Big(  {N\ov 4 \pi \l} + { \pi \l \ov 6 N }  +...\Big)  \, e^{\pi \sqrt{2\l} } 
=  \Big(1+     { \pi   \ov 12 }\,   { g^2_\str \ov T}\ +...\Big)\, W_1 \ , \qquad\qquad 
W_1 = { \sqrt T \ov \sqrt{ 2 \pi} \,g_\str}  \, e^{2 \pi T } \ .
\ee
Here $W_1$ is the leading disk  term in $\langle \W\rangle$ in the ABJM theory given by \rf{100}  (with $\bar \G_1=0$  according  to  \rf{01}).
Thus, to this order, the   genus expansion  in the  ABJM case 
  has the same universal structure  as  \rf{445},\rf{446} in the SYM case. 
  Remarkably, the coefficient  ${\pi \ov 12}$ of the ``one-handle" term $g^2_s\ov T$ 
  is precisely  the same as in the  SYM  case \rf{446}, calling for some universal  explanation. 
  It would be interesting to see    if  the prefactor in \rf{447} exponentiates as in \rf{446} 
  (e.g. using the results of  \cite{Klemm:2012ii})
  and also if there is a D2-brane description of this similar to the one in the SYM case  discussed above (cf. 
  \cite{Drukker:2008zx,Cookmeyer:2016dln}).


\section{Concluding remarks \la{s6} }
\def\theequation{6.\arabic{equation}}
\setcounter{equation}{0}
As was noted below \rf{2211},  the coefficient of the UV   divergent term in \rf{317} is, in fact,  
the same  for all minimal surfaces with disk topology, and thus  the dependence of the string partition 
function  on the   scale $\rR$  or effective tension $T$  through the $\sqrt T$     factor 
in \rf{3077} should be universal.  

A  check   of the  universality of the  prefactor in  \rf{100}  is that  it applies also to the 
 circular WL in the $k$-fundamental representation  dual to a minimal  surface  ending on 
$k$-wrapped  circle  at the boundary of AdS$_5$.  In this   case  the  classical 
action is $I_{\rm cl}= - k \sql $  but  the Euler  number of the minimal 
surface is still equal to 1  
   \ci{Forini:2017whz} so that the coefficient $\zeta_{\rm tot}(0)$ in \rf{317},\rf{3077} 
is  also 1  and thus the disk partition function is
  $\langle \W\rangle \sim {\sqrt T\ov g_\str}\,  e^{2 \pi k T}$.
This is  consistent   with  the  
  SYM (localization) result in the $k$-fundamental case \ci{Drukker:2005kx,Drukker:2005cu,Pestun:2007rz}  given  by the $k=1$  expression with $\sql \to k \sql$. The    
 overall $k$-dependent  constant that  should come from $\bar \G_1$  in 
\rf{100}  still remains  to be explained, despite several 
earlier attempts in \ci{Kruczenski:2008zk,Buchbinder:2014nia,Bergamin:2015vxa,Forini:2017whz}.

The universality of \rf{100}   implies, in particular, that the  prefactor $ \sqrt T \ov g_{\rm s}$ 
should  cancel in the ratio of  expectation values of similar Wilson loops.
In particular,  this    applies 
to   the  case of   $1\ov 4$  BPS     latitude  WL 
parametrized   by  an  angle $\theta_0$. 
Matching with the gauge  theory prediction  for the ratio  of the latitude WL   and  simple  circular WL    was checked 
 in the SYM  case in  \cite{Cagnazzo:2017sny,Medina-Rincon:2018wjs}  and  in the ABJM   case in  \cite{Medina-Rincon:2019bcc,David:2019lhr}.

Let us note that  \rf{100}  actually   requires  a  generalization in  special   cases  when  there are 
 0-modes  in  the  internal  (non-AdS)   directions  of ${\rm AdS}_n  \times M^{10-n}$ 
 space,   each  producing extra factor of $\sqrt  T$ (cf.  \rf{303}).
This is what  happens  in  the  case of the $1\ov 4$  supersymmetric   ($\theta_0={\pi \ov 2}$)   latitude WL
 discussed in \ci{Zarembo:2002an,Medina-Rincon:2018wjs}  where   we then 
  get  for the disk partition function 
 \be \la{500}
 \langle \W_{1\ov 4} \rangle  \sim {\sqrt T \ov g_\str}\,  ( \sqrt T)^3 \sim N 
 \ .\ee 
  Here all   $\l$-dependence  cancels out  
 and the finite  proportionality  constant  should be  equal to 1, i.e.  
   $N^{-1} \langle \W_{1\ov 4}  \rangle =1$, 
 in agreement with  \ci{Zarembo:2002an}.

A similar  remark  applies to the case of the ${1\ov 6}$   (bosonic) 
  BPS  WL  \cite{Drukker:2008zx}  in  the ABJM  theory. 
According to \ci{Drukker:2010nc}  here   we get instead of $\langle \W \rangle$ for the $\ha$ BPS WL 
 in 
   \rf{55} (cf. footnotes \ref{fo1}, \ref{fo2}   and eq.\rf{447})
\be \la{588}
 \langle \W_{1\ov 6} \rangle = { 1 \ov g_{_{\rm CS}}}  \langle \WW_{1\ov 6}\rangle_{_{\rm loc} }
 =  i e^{i\pi \lambda}    { N \ov 4 \pi \l }   \,  \tfrac{1}{ 2} \sqrt{2 \l} \   e^{\pi  \sqrt{2 \l} } +... \ . 
\ee
As  was argued in \cite{Drukker:2008zx}  (see also \cite{Rey:2008bh}),   here  the minimal surface 
solution is smeared over  $S^2=CP^1$ in $CP^3$   so there are two scalar 0-modes. 
This explains  the extra factor $ \ha   { \sqrt{2 \l}} = (\sqrt T )^2$ in \rf{588} compared to  $\langle \W \rangle$ in   \rf{55}
 \ci{Drukker:2010nc}.
More generally,  for  contributions  from 
 each  genus $p$ one finds   \ci{Marino:2009jd,Drukker:2010nc,Klemm:2012ii}   that  the ratio of the   ${1\ov 6}$    and ${1\ov 2}$ 
  BPS  WL's  is given by this universal  $(\sqrt T )^2$ term (ignoring phase factors)  
\be \la{599}
{\langle \W_{1\ov 6} \rangle_p \ov \langle \W 
\rangle_p} =  ( \sqrt  T)^2  + \OO ( T^{-1}) \ .\ee
It would be interesting to match the precise numerical coefficient in the ratio between the ${1 \ov 6}$ BPS and ${1 \ov 2}$ BPS Wilson loops by carefully fixing the normalization of the two zero modes on the string side. 

Finally, let us note that while in this paper we focused on the case of 4d and 3d gauge theories, as explained in section \ref{s2} our results
 also apply to string theory in ${\rm AdS}_3 \times   S^3\times T^4$ with RR flux.  
 This   case corresponds to $n=3$ in \rf{105} (cf. \rf{100},\rf{1000x}), i.e. 
 $\langle \W\rangle =Z_{\rm str} = \frac{1}{g_\str}\,\sqrt{T}\,e^{2\pi T}+\ldots$. 
 It would be interesting to see if this string-theory  prediction   can be matched to localization calculations 
 for Wilson loops in 2d supersymmetric gauge theory (cf. \cite{Sax:2014mea,Doroud:2012xw}).

  \iffa 
(1) straight string:  here  we expect $Z=1$, but then  we have  puzzle with $ 1/g_\str$   proposal --
dependence on $N$?  NO,  $ 1/N tr exp =1$,  so overall $N$ is ok. 
Here  regularized  volume of \adt   is  zero,  so all 0-modes   are not normalizable 
   so we could think that this  explains $Z=1$   but $ 1/g_\str$   factor is still there. 
The point may be that  SYM on $R^4$  vs localization is 
not well defined so we are to look at $R \times S^3$   or strip geometry of \adt.  
\fi

\section*{Acknowledgments}
We  are grateful to  K. Zarembo  for  useful discussions and  remarks on the draft. 
We also thank  M. Beccaria,    P.  Di Vecchia,  N. Drukker,   M. Kruczenski, J. Maldacena,  R. Roiban   and  E. Vescovi  for  helpful  comments. 
The work of  S.G. is supported in part by the US NSF under Grants No.~PHY-1620542 and PHY-1914860.
A.T. acknowledges the support of the 
 STFC grants ST/P000762/1  and ST/T000791/1.

\iffa
 
Measure in path integral in general is not uniquely defined. 
Counterterms  may   reflect  this ambiguity.  Sometimes symmetries fix the measure. 
For example, $\sqrt G$ in sigma-model measure cancels   quadratic divergences;  alternative is to add corresponding volume  counterterm. 
In string context these   may be $\int R$   and boundary counterterms like $\int K$, length, etc.

Just to add a comment that I think was already mentioned before: this normalization factor in Z that we are after, sqrt{T/2pi}, should be universal and not depending on details of {\rm AdS}_2 spectra etc. It should be the same factor for all solutions with disk topology. This is because the calculations taking the ratio of Wilson loops have been shown to work, so this factor just has to cancel out in the ratio.
The pattern of cancellation I recalled in previous email works the same way for any solution, not just AdS2, if we note that in general we should have for the sum of boson and fermion masses sum (m_b^2-m_f^2) = -R, where R is 2d curvature. I think this sum rule should apply in general for semiclassical expansion around any solution (the paper by Forini et al also shows this explicitly).
So in the end this power of T in the Z normalization should just depend on \chi...which smells again as similar to an explanation in terms of zero mode counting.
So maybe we'd better not get into convoluted explanations that have to do with the details of the fermion's Green's function in AdS2...
\fi 

\appendix
\section{Comments on  tension dependence of the string partition function \la{apa}}

In section \ref{s3} we discussed how to  explain   the  prefactor $\sqrt T$  in the one-loop string partition function  \rf{100} 
starting with  the  string action \rf{003}  and using the  static gauge  expression \rf{b2}.
We emphasized that the result is sensitive to the choice of the path integral measure, i.e. the definition of the quantum theory 
(which,  in general,  is not unique, unless  completely fixed by  symmetry  requirements or extra consistency conditions). 
In the Appendices below  we shall   discuss  some other  approaches  to derive this prefactor, which again involve  certain assumptions about 
the measure or regularization procedure. 

\subsection{$T$-derivative of the partition function in static gauge \la{a1}}
\def\theequation{A.\arabic{equation}}
\setcounter{equation}{0}
 
Suppose we    start  with  the string action \rf{007}   in terms of the rescaled  (dimensionless) 
  coordinates    so that there is an explicit factor of  the effective string tension 
  $T$  in front of the action  with  the induced \adt metric  having radius 1. 
  Then we  would get the same result as  in \rf{3077}    if we assume that the norm or the measure  is defined so that  the resulting 
  one-loop correction  from a single scalar has   the form 
  $\Gamma_1 = \ha \log \det \hat \Delta$ where $    \hat \Delta=  T^{-1} \Delta $.\foot{Explicitly,   
  $( x, \hat \Delta  x) = T  \int d^2 \s \sqrt g\,   x \hat \Delta x =   \int d^2 \s \sqrt g\,    x \Delta  x$, 
  where $ x$ are rescaled fluctuations.
  In general, one can of course move   $T$-dependence from the action to the measure  by a field redefinition 
   (taking into account   the resulting regularized Jacobian of the transformation). 
    If the path integral  measure     is  $ \prod_\s {\mu \, \ov \sqrt{ 2 \pi}}   dx(\s)  $
 and the action is  simply 
  $\ha  \int  d^2 \s \sqrt g\,     x \Delta  x$
then $Z= (\prod_n { \l_n \ov \mu ^2})^{-1/2} \sim     \mu ^{\zeta(0) } $
where $\l_n$  are the eigenvalues of $\Delta$ 
  \cite{Hawking:1976ja}. 
If  $\zeta(0)$  is non-zero  the  result   is thus  sensitive to  the definition 
of the measure.}
    Indeed, using  the  $\zeta$-function regularization 
  with   $\zeta(0)$   being   the regularized   total number   of  eigenvalues we get 
  $ \ha   \log \det (  T^{-1} \Delta )   = \ha  \zeta(0) \log ( T^{-1}) + ... = -   \zeta(0)   \log \sqrt T + ...$. This   leads  again 
   to \rf{3077} 
    once we use that  the total  value  of $\zeta(0) $   corresponding to the static-gauge partition function \rf{3} is 
        $\zeta_{\rm tot}(0)=1$ (see \rf{2211},\rf{1200}). 
  
  Another  way  to obtain  the same result  (which will be again based on a  particular  choice of a  regularization prescription) 
  is to find   the dependence of the string  partition function on the tension 
   by first   computing its derivative over $T$. This is  closely related  to 
    the argument   appearing in the context of  the 
     ``soft dilaton theorem"   \cite{Tseytlin:1988ne}, see  section \ref{sc3}.
     
  Let us    assume that the   tension  dependence  of the string partition function 
   may come only  from the 
factor  of $T$  in the  string  action \rf{007} in the  static gauge (i.e.  in  the action for  the  ``physical"  fluctuations  whose determinants   are present in \rf{b2}), i.e. the measure is  defined so that it does not depend on $T$. 
 For example, for   a single scalar field  
\be  \la{si}
&&Z= \int [dx]   \exp ( -   I ), \ \qquad    I  =  \ha   T \int  d^2 \s\sqrt g\,    x  \Delta_{(m^2)}    x  \ ,\qquad 
 \ \ \Delta_{(m^2)} = - \nabla^2 + m^2\ , \\ 
&&  \la{37} 
 T { \del  \ov \del  T } \log Z  = - \langle I  \rangle \ , \ \ \ \ \ \ \ \ 
  \langle I  \rangle =  \int d^2 \s\sqrt g\, [  \Delta   \rG_{(m^2)}(\s,\s')]_{\s=\s'} =  \int d^2 \s\sqrt g\, \delta_{(m^2)}(\s,\s) ,
\ee
where $\langle I \rangle = Z^{-1}  \int [dx] \,    I \,   \exp ( -   I )$, \ \ 
$\rG_{(m^2)}(\s,\s') =  \langle \s|  \Delta^{-1}_{(m^2)} | \s'\rangle $ is the Green's function  and 
$\delta_{(m^2)}(\s,\s)$  is a  regularized value of the 
 bosonic delta-function at  the coinciding points. Let us   use the heat-kernel cutoff, i.e. assume that  
  \be  \la{379}
 \delta_{(m^2)}(\s,\s) =  \langle \s|  e^{- \eps\,  \Delta^{-1}_{(m^2)}}  | \s\rangle 
 =\te   {1\ov 4 \pi} \big[  \Lambda^2  + {1\ov 6}R^{(2)}   -  m^2 \big] \ , \ \ \  \  \ \ \  \eps\equiv  \Lambda^{-2}\to 0  \ .\ee
The expectation value of the action   corresponding to the  full   static-gauge  expression \rf{b2} is then
\begin{align} 
 \la{38}
 \langle I \rangle  &= \ha  \int d^2 \s\sqrt g\, \Big\{ (n-2)   [  \Delta_{(2)}    \rG_{(2)}  (\s,\s')]_{\s=\s'}  
 + (10-n)   [  \Delta_{(0)}    \rG_{(0)}  (\s,\s')]_{\s=\s'}  
 \no\\
  & \qquad \qquad   \qquad \qquad    -  (2n-2)     [  \D^f_{(1)}    \rG^f_{(1)}  (\s,\s')]_{\s=\s'} 
 -  (10-2n)     [  \D^f _{(0)}    \rG^f_{(0)}  (\s,\s')]_{\s=\s'} 
  \Big\} \ ,   \\
 &= \ha  \int d^2 \s\sqrt g\, \Big\{  (n-2)  \delta_{(2)}(\s,\s)  + (10-n)   \delta_{(0)} (\s,\s)
 -  (2n-2)   \delta^f_{(1)} (\s,\s)    - (10-2n)   \delta^f_{(0)} (\s,\s)   \Big\} \ , \no 
\end{align}    
where   $\D^f$  is the fermionic   1st order operator    and  $\rG^f$  and $ \delta^f$  stand   for the 
corresponding  Green's function and $\delta$-function. 

A  key  next step is to assume a special 
  ``2d supersymmetric"  regularization  in which the bosonic   and fermionic  Green's functions and thus 
  the corresponding regularized   delta-functions are related to each other  as\foot{This is
    an effective consequence  of 
    the fact  that the 2d supersymmetric Ward identity  
    (cf. \ci{Bardeen:1984hm,Inami:1985di})  relates 
     a fermion of   mass $ m > 0$  to a  boson of  mass $m^2-m$
     (e.g.  in a special regularization   the  trace of the Green's function  for a single 2d fermion  $ \rG^f_{(m)}$
    is  related to   $ 2 m\,  \rG_{(m^2-m)}  $ \ci{Inami:1985di}).
        In   the present case  we have 
     half  of  the  massive  fermions   with   mass  $m=1 $ and  the other half -- with mass $m=-1$.
Alternatively,  one may use a  particular representation for  $ \rG^f_{(m)}$ (for $m >0$) as 
$\rG^f_{(m)} (\s,\s')  = [( i \gamma^a  \del_a   + m)  \rG_{(m^2-m)}   ] S (\s,\s')$ 
      implying that    one has 
       $  \D^f_{(m)}    \rG^f_{(1)}  (\s,\s') =  \delta_{(m^2 -m )}(\s,\s')\, S (\s,\s')$. 
      }
\be \la{39}
  \delta^f_{(m)} (\s,\s) = \ha \Big[     \delta_{(m^2 -m )}(\s,\s) +    \delta_{(m^2 + m )}(\s,\s)\Big] \ .
\ee
Then \rf{38}  reduces  simply to 
\be \la{391}
\langle I \rangle  = \ha  \int d^2 \s\sqrt g\, \Big[  \delta_{(0)}(\s,\s) - \delta_{(2)}(\s,\s)  \Big] 
 =\te  { 1 \ov 2} \times  {1\ov 2 \pi}  \times V_{\rm {\rm AdS}_2} = - {1\ov 2} \ ,  
\ee
where    we used \rf{379}.\foot{Let us note that 
the  use of \rf{39}  may be interpreted as  a specific regularization  prescription   for  the fermions   
which is 
different from the heat-kernel or $\zeta$-function  one applied 
to  the squared fermionic operator $\Delta^f_{(m^2)}= (\D^f_{(m)})^2 = -\nabla^2 + {1\ov 4} R^{(2)}  + m^2  $  in \rf{b2},\rf{2}. 
Indeed, if we
 assume  that  $ \delta^f_{(m)} (\s,\s)$ in \rf{39}  is defined as in \rf{379}, i.e. 
$\delta^f_{(m)} (\s,\s) =  \langle \s| e^{- \eps \Delta_f} | \s \rangle
 = {1\ov 4\pi} \big( \Lambda^2  +   {1\ov 6} R^{(2)} - {1\ov 4} R^{(2)}  - m^2 \big) $
 then    we    find   that 
  $ \langle I \rangle = + \ha$   which is 
   consistent with the $\zeta_\tot (0)=1$ value in \rf{2211},\rf{1200},  i.e. $ Z \sim \prod [\det ( T \Delta)]^{-1/2}   \sim T^{-1/2}$.
 In this regularization  the  l.h.s. of  \rf{39}  is 
 $ 2 ( \Lambda^2  +   {1\ov 6} R^{(2)})  - {1\ov 2} R^{(2)}  - 2 m^2  $  while the 
 r.h.s. is    $ 2 ( \Lambda^2  +   {1\ov 6} R^{(2)})  - {1\ov 2} R^{(2)}  - 2 m^2  $ 
 so the difference  $ - {1\ov 2} R^{(2)}$  may be attributed   to the presence of $- {1\ov 4} R^{(2)} $ term in the 
 squared  fermionic operator   which is  thus effectively omitted in the prescription  \rf{39}.
  } 
  
  The relation $ T { \del  \ov \del  T } \log Z = - \langle I  \rangle$ in \rf{37}   implies  once 
again that
\be \la{393} 
Z \ \sim \ \sqrt T   \ .  \ee
Let us note that the  result  for the expectation value of the action \rf{391}  should be more
 universal  than a particular   prescription used  above.
The integrand  in \rf{391}  should be  in general $ \delta_{(0)}(\s,\s) - \delta_{(2)}(\s,\s) \to -  {1 \ov 4 \pi} R^{(2)}$.
In the case of a more general topology  of a 
 disk with $p$ handles  with  the Euler number $\chi=1-2p$ we should  then  find  that 
\be \la{399}
\langle I \rangle  = - \ha \chi \ , \qquad \qquad    Z \sim  (\sqrt T)^\chi   \ ,  \ee
which is  in agreement   with \rf{1000},\rf{445}.

\subsection{$T$-dependence from zero modes in conformal gauge \la{aa2}}
The  conformal gauge expression for the string partition function  contains, in addition to  the ratio of determinants in $Z= e^{-\G_1}$ in \rf{b2}, also  an extra  factor \ci{Fradkin:1982ge,Drukker:2000ep} 
\be \la{3003}
Z_{\rm  c }= \Omega^{-1} \Big[  {\det' \Delta_{\rm gh} \ov \det \Delta_{\rm long}} \Big]^{1/2} \ . \ee
Here   $\Omega$ is the $SL(2,R)$   Mobius group volume. 
  The  2-derivative  ghost operator $\Delta_{\rm gh\, ab}$ and the  operator on  the two 
``longitudinal" fluctuations  $\Delta_{\rm long\, ab}= - (\nabla^2)_{ab}  -   \ha R^{(2)} g_{ab}$
have the same structure  (and  the same   ``mixed"    boundary conditions)
so   their  non-zero-mode  contributions  should effectively   cancel each other. 
\iffa \foot{In the case of  flat  target 
space starting with the  longitudinal directions $x^m$ and setting $x^m = \xi^a \del_a \xx^m$  where $\xx^m$  is classical solution 
$L= h^{ab } \del_a x^m \del_b x^m$   where $h_{ab} \del \xx^m \del_b \xx^m$ is  induced metric 
gives same as the ghost operator   on $\xi^a$ -- conformal factor cancels so the zero modes are CKV:
in terms of $\xi'^a = e^\r \xi^a$, $\del_a  \xx^m = e^\r \delta^m-a$. 
In curved space  we need to rotate coordinates to tangent space; this is included in induced metric and   after that longitudinal and ghosts are related.
}  \fi 
The integral over the collective coordinates of the three    0-modes of $ \Delta_{\rm long}$   (or conformal Killing vectors) 
which  is implicit in \rf{3003}
should   cancel   against the Mobius    volume factor.  As  a  result, 
    one may assume   that $Z_{\rm  c }$  in \rf{3003} 
is effectively equal to  1, thus getting back to the static gauge partition  function expression \rf{b2}. 

However,   this   depends on the  definition of path integral measure. 
An alternative   possibility  compared to the one  in the static gauge discussed  in 
the main text and section \ref{a1} 
 is  to assume  that   in the conformal gauge the  measure is   defined so   that   the 
   path integral  over  all  non-zero modes   does  not produce any $T$-dependent factor,  
 while  the presence of the $\sqrt T$ factor   is \rf{100}  is  due to the  normalization of the 0-modes, i.e.
 of the collective coordinate integral implicit  in \rf{3003}.\foot{As was already mentioned above, this corresponds to a    specific choice of the measure factors  implying   that the 
    normalization of the gaussian  path integral  is 
     1, i.e. $\int [dx]  \exp[ - { 1 \ov 2 \hbar} (x,  x) ] =1$,   with   $[dx]= \prod_\s
   {dx(\s) \ov \sqrt{2 \pi \hbar}}  $.
Then the  factor of string tension $T= \hbar^{-1}$   should appear  not only in the action but also in  the measure 
 so that it  cancels out in the integrals for all  modes  with non-zero eigenvalues.} 
 Each   bosonic 0-mode   absent in the fluctuation  action  then  contributes a  measure factor  
 $\sim  ({T\ov 2\pi})^{1/2}, $   leading to 
 \be   Z \ \  \sim \ \    (\sqrt T)^{n_0}  \  ,  
  \la{303} \ee
  where $n_0$ is the total number of the 0-modes. 
 
  Equivalently, this  result will follow    assuming  that one  uses a regularization (e.g., dimensional one) in which 
  the delta-functions at coinciding points  vanish,  $\delta^{(2)}(\s,\s)=0$  and thus the factors of $T$  in the measure and 
  in front of the action  do not contribute, cf. \rf{37}, apart from the 0-mode  contribution.
\iffa
 Let us note that  depending on regularization
 of the contribution of  all other modes 
   this  approach may or may not be equivalent to the  one discussed earlier. 
    Indeed, if it happens that the total  (regularized)  number  of all modes  vanishes, i.e. 
    $B_2= \zeta(0) + n_0 =0$, which is the case  of a  UV finite theory,  then 
 the  regularized number of non-zero modes is    $\zeta(0) = - n_0$, and thus 
 the result  $(T^{1/2})^{-\zeta(0)}$ 
   (with all determinants  being restricted to non-zero modes)  is the same  as   \rf{303}. 
Since the dependence on $T$   coming  both from the measure and   the  action is effectively controlled by 
the delta-function at  coinciding points 
$\delta^{(2)}(\s,\s)$   (given  by \rf{379}  in the heat-kernel regularization)  one may also   achieve the equivalence by assuming a regularization where $\delta^{(2)}(\s,\s)=0$.
 At the  same  time, if the theory  is   not necessarily UV finite   but 
 contributions of   non-zero bosonic and fermionic modes mutually cancel (as 
 happens,  e.g.,  in supersymmetric theories  in  an  instanton  background  \ci{Shifman:2012zz}) 
 then   the total $\zeta(0) = 0$ and  thus the    coefficient of   log UV divergence 
 is  given just by the  number of the   zero modes,  $B_2 = n_0$.
 \fi
 \iffa 
 Then  the  two approaches   are no longer equivalent.  The dependence on  coupling 
 or $T$    in the  latter  supersymmetric case is  actually ambiguous  in what concerns the fermion 0-mode measure dependence on coupling;  one natural 
 prescription    gives (assuming fermions are Majorana)  $T^{n_b -{1\ov 2} n_f}$  (see  \ci{Shifman:2012zz}).\foot{In the 4d gauge theory context  the role of $T$  is played by  the  coefficient $1/g^2$
 in front of the SYM action  with   the norm of 0-modes  proportional to 
 square root of the value   of the classical action  or $1/g$.
  Unambiguously defined  observables  are non-zero  correlation functions   with external fermionic legs (see discussion in 
 \ci{Novikov:1983uc}).}  At the same time the  UV cutoff factor in the one-loop partition function  is  $
 \Lambda^{B_2} = \Lambda^{n_b - n_f}$. 
\fi 
It is useful  to   recall that  in the  familiar    case  of  the  open strings with free ends where 
the  bosonic  coordinates $x^m$ ($m=1, ..., D$)   are subject  to the Neumann  boundary conditions  
  one finds 
  $D$   constant zero modes  and thus  an  overall   factor of $T^{D/2}$ in the disk path integral. The same  
  result  can be   found also using $T {\del \ov \del T}$  argument   by using    that the  delta-function appearing 
  in \rf{37}  is the   ``projected" one, i.e.    $\delta^{(2)}(\s,\s)$  (set to 0)  minus the trace of the  projector to the 0-mode subspace  (see, e.g.,    \cite{Tseytlin:1988ne,RandjbarDaemi:1987aj}).

 In the  present WL  case of    path integral  with  the Dirichlet-type  (or 
 fixed-contour)   boundary conditions   one could expect to have  no 
0-modes. 
However,  as the two ``longitudinal"   string coordinates are subject to   ``mixed"  Dirichlet/Neumann   b.c. 
\ci{Durhuus:1981ad,Fradkin:1982ge,Alvarez:1982zi,Luckock:1989mv} (motivated by  the requirement of
 preservation of the  reparametrization  invariance 
of the  boundary contour)
  there is, in particular,   a special 
 0-mode corresponding to a constant shift of a point on the boundary circle. 
There are, in fact,    two   more 0-modes of the longitudinal  operator (see Appendix  \ref{b1}). 
   As already mentioned above, these  three  bosonic 
  0-modes  are direct counterparts of the conformal Killing vectors   associated to 
   the $SL(2,R)$ Mobius  symmetry on the disk
 surviving in the conformal gauge.

 Thus if   the path integral   measure is normalized  so that the integral over non-zero modes does not 
 produce any  $T$-dependence 
 we then  get a   factor $ Z \sim  (\sqrt T)^3$  associated to  the  $n_0=3$   ``longitudinal"   0-modes on the disk. 
 To  reduce the   effective number of 0-modes to $n_0=1$ (required to match  the $\sqrt T $ factor  in \rf{100})
   one may contemplate the following possibilities:
   
  (i)  assume   that 2   
 longitudinal    0-modes   are  lifted due to some  boundary  contributions to  the string action 
    leaving only one    translational 
  mode (corresponding to a 
  constant  shift on the     boundary circle);\foot{This,  at first, may  look 
   unnatural as  then  we would  not have a
  cancellation between the integral over the corresponding 
  collective coordinates and the Mobius volume factor  in the path integral.
  Yet,  that may not be a problem as  the  Mobius volume  on the disk  
  may  be regularized to a finite value 
 \ci{Tseytlin:1987ww,Liu:1987nz}  (similarly to how this is done for   the \adt volume).}

 (ii)  assume  that the  GS  fermion  contribution effectively  conspires to mimic the  NSR 
  contribution on the disk with $n_f=2$ fermionic  super-Mobius 0-modes,\foot{To compare, 
 in the case of the one-loop  instanton partition function in  super YM  theory   (see \ci{Novikov:1983uc,Shifman:2012zz})
 the  contributions of 
 all non-zero modes cancel (i.e. $\zeta_\tot(0) =0$)   and as a result  the UV cutoff dependence (and thus one-loop beta function) 
  is controlled  just by the  0-modes --   the total Seeley   coefficient  is 
 $B_4 = \zeta_\tot(0) + n_{\tot}=   n_b - \ha n_f$. 
 At the same time,   the dependence on the inverse   gauge coupling 
 $1/g^2_{_{\rm YM}}$  (which is the analog of string tension $T$  in our  case)  is  controlled by 
 the coefficient $n_b-n_f$. 
Note, however,  that 
 the prescription for $g_{_{\rm YM}}$ dependence becomes unambiguous only in physical correlation functions 
 with external fermionic legs saturating  the  fermionic 0-mode integral  \ci{Shifman:2012zz}.}
   producing the 
   effective  number   $n_0= n_b-n_f= 3-2=1$.
   A relation to
   the   NSR  formulation with  manifest 2d supersymmetry  may   of course  be 
   expected   and was mentioned already in the discussion of the static gauge approach  above. 
    Note also that the super-Mobius volume
    is finite \ci{Andreev:1988cb}   so it is  not  necessary to cancel it explicitly.
  
\iffa 
One problem with this  suggestion  is that 
 it is not clear  why the path integral would not  just vanish due to the   presence of  fermionic 0-modes.
Also, the  fermionic measure normalization does not have extra   factor of $1\ov \sqrt{2  \pi}$ (required for bosonic gaussian integral)    so we will not reproduce that way  the $( 2 \pi)^{-1/2}$ factor  in  $ \sqrt{\frac{ T}{ 2 \pi}}$. 
Furthermore,    it is not clear  how these longitudinal "fermionic  0-mode"
contributions   should appear  in the GS  superstring framework.  
\fi 



\def \aa  {{\rm a }}
\def \bb  {{\rm b }}

\section{Conformal Killing vectors  as longitudinal zero modes  \la{b1} }
\def\theequation{B.\arabic{equation}}
\setcounter{equation}{0}

Here we shall 
record   the expressions for the  conformal gauge ghost zero-modes or conformal Killing vectors 
  (CKV) on a  flat disk $D^2$ and on a   euclidean  hyperbolic space 
  $H^2={\rm AdS}_2 $  with the metric\foot{Alternatively,  for the \adt   metric we have 
$ds^2 = ( \sinh^2 s)^{-1} ( ds^2 + d \p^2), \ \   r= e^{-s}$.
Another form of the \adt  metric   that follows  from  ${\rm AdS}_3$ 
   metric $ds^2 = z^{-2} ( d\rr^2 + \rr^2 d \p^2 + dz^2)$ with $z= \sqrt{1-\rr^2}$ is 
$ds^2 = { d \rr^2 \ov (1- \rr^2)^2 }  +  {\rr^2 d\p^2 \ov 1-\rr^2} $ 
is related to the above one  via $\rr= { 1 \ov \cosh s} = { 2 r \ov 1+ r^2} $.
}
\be 
ds^2 = e^{2 \rho} ( dr^2 + r^2  d\p^2) \ ,  \ \ \ \  \ \ \     (e^{2 \rho})_{D^2} =1 \ , \ \  \ \ \ \   (e^{2 \rho})_{H^2}   = {4\ov (1-r^2)^2} \ . 
\ee
The  CKV  are also  the zero-modes  of the longitudinal   Laplacian in \rf{3003} 
which is equivalent to the  2nd derivative  conformal ghost operator \ci{Fradkin:1982ge}.
The defining   relation $\na_a \xi_b + \na_b \xi_a - g_{ab} \na^c \xi_c=0 $ 
does not depend on  the conformal factor $\rho$ when written in terms  of the contravariant  components 
$\xi^a$: 
 $\del_a \xi^b + \del_b \xi^b - \delta_{ab} \del_c \xi^c=0 $.
The  expressions for the three    Killing vectors  $\xi^a$ corresponding to the $SL(2,R)$  
 transformations on the plane are (here $\aa,\bb_1, \bb_2$ are    real parameters) 
\be 
&&z'= e^{i \aa}{   z + \bb \ov 1 + \bb^* z }, \qquad   \bb= \bb_1 + i \bb_2, \qquad 
\qquad \delta z =\xi^1 + i \xi^2   =  \bb + i \aa\, z - \bb^* z^2, \qquad    z= r e^{i \p}\ , \\
&& \xi^1 = \bb_1 - \aa\, r \sin \p - r^2 ( \bb_1  \cos 2 \p  + \bb_2 \sin 2\p)\ , \quad 
\xi^2= b_2  + \aa\, r \cos \p - r^2 ( - \bb_2 \cos 2\p + \bb_1  \sin 2\p) \no \\
&& \xi^r = \cos \p\  \xi^1 + \sin \p\ \xi^2 ,\ \ \ \  \xi^\p = r^{-1}  ( - \sin \p\ \xi^1 + \cos \p\ \xi^2)\,  \no \\
&&
 \xi^r =       (1-r^2) ( \bb_1 \cos \p + \bb_2 \sin \p)\ , \qquad \qquad 
\xi^\p = \aa + (r+ r^{-1}  )  ( \bb_2 \cos \p - \bb_1 \sin \p) \la{a2} \ . 
\ee 
 Then  the  standard  conformal Killing  vectors  on the  disk satisfy  mixed   boundary conditions:
 $\xi^r=0 $ (normal component)  and $\del_r \xi^\p=0$  (normal derivative of tangential component) 
  vanish at  the $r=1$  boundary. 
   Once we  consider a metric with a  non-trivial  conformal factor these conditions are  modified to:
  \be \la{a3} 
  g_{ab}= n_a  n_b + t_a t_b \ , \ \ \qquad   \xi_n \big|_\del =0 , \ \ \ \ \   (\del_n - K)  \xi_t \big|_\del  =0 \ , \ \ \  \ \ \   K= \na_a  n^a \ . \ee
  The mixed 
    boundary conditions were discussed in \ci{Durhuus:1981ad} and \ci{Fradkin:1982ge}. The 
   condition 
  $  (\del_n - K)  \xi_t \big|_\del  =0$ was used in \ci{Luckock:1989mv} (and implicitly  in \ci{Alvarez:1982zi}).
  
  The $r, \p$ components of $n_a$   and $t_a$ are:   $n_a = e^\r \{1,0\}, \ \ t_a = e^\r \{0, r\} $  so 
  that 
  \be 
&&   \xi_n = n_a \xi^a = e^\r  \xi^r \ , \qquad  
   \xi_t = t_a \xi^a  = r  e^\r  \xi^\p \ , \\ 
 &&   \ \ \del_n = e^{-\r} \del_r \ , \ \ \ \   
   \ \ K= e^{-2 \r} r^{-1} \del_r  ( r e^\r) \ , \ \ \   ( \del_n -K) \xi_t= r \del_r \xi^\p \ . 
   \ee
   Note that  for a  flat disk $K=r^{-1} $ and $\chi = {1\ov 4 \pi} ( \int R +  2 \int_\del  K ) =1$.
   
   Thus $\xi^\p$ in \rf{a2}   satisfies  $ (\del_n - K)  \xi_t \big|_\del  =0 $  at $r=1$ 
   but  there is  an issue with $ \xi_n \big|_\del =0$: \ \ 
   $e^\r  \xi^r =    {2 \ov 1-r^2} \times (1-r^2) ( \bb_1 \cos \p + \bb_2 \sin \p)$  so $\xi_n$ is  a  non-zero function at the boundary. 
   This suggests that  either we should   set $\bb_1,\bb_2 =0$ or the  boundary condition $ \xi_n \big|_\del =0$ 
   is to be modified.
     One option is to define it 
   with the flat metric  as in (2.15) in \ci{Fradkin:1982ge}:  $\td n_a  \xi^a |_\del =0$  where  $\td n_a$ is the normal in flat metric. 
    This condition just says that the boundary condition 
   $x^m|_\del  = c^m(\p)$   should be preserved under diffeomorphisms  up to a boundary reparametrization, so 
   $\delta x^m = \xi^a \del_a x^m $ should   vanish at the boundary   for $\xi^a$ along  the normal  direction
   (the definition of normal formally depends on the metric, but here all we need is  $\xi^r\big|_\del =0$). 
   

The norms of CKV   depend on the conformal factor:\foot{The definition of the norm for  the diffeomorphism vectors 
 via  $|\xi|^2 = \int d^2 z \sqrt g\, g_{ab}\,  \xi^a \xi^b $ is a  natural one; while 
 it  involves  the conformal factor,     it is the  conformal factor dependence in the 
 corresponding   determinants  that should cancel in  the critical dimension. This definition is different from the one used in 
   \ci{Alvarez:1982zi}  but agrees   with the one of  \cite{Moore:1985ix,Luckock:1989mv}. 
   For a discussion of the  freedom in the choice of the path integral measure see also \cite{Weisberger:1986qd}.}
\be \la{a11} 
|\xi|^2 = \int d^2 z \sqrt g\, g_{ab}\,  \xi^a \xi^b =  \int^{2\pi}_0 d\p \int^1_0 dr \, r \,   e^{4 \rho} \,   \xi^a \xi^a  \ .\ee
For the three   CKV  proportional to $\aa, \bb_1, \bb_2$ in \rf{a2}  we have  ($\xi= \{ \xi^r, \xi^\p\}$):
$\
\xi_{(\aa) }= \aa \big\{0, 1\big\} ,$  \ \  $  \xi_{(\bb_1)}=   \bb_1  \big\{ (1-r^2) \cos \p, \ - ( r + {r^{-1} } ) \sin \p\big\}  ,$ \ \ 
$ \xi_{(\bb_2)}=   \bb_2  \big\{(1-r^2) \sin \p, \  ( r + {r^{-1}} ) \cos \p\big\} $.
Thus for $\xi^a \xi^a $ in \rf{a11}   we get:\ 
 $\xi_{(\aa)} \cdot \xi_{(\aa)} = \aa^2 r^2,$ \ \ $
\xi_{(\bb_1)} \cdot \xi_{(\bb_1)} = \bb_1^2 [ 2  (1+ r^4)  -   2 r^2 \cos 2\p] ,$ \ \ $   \xi_{(\bb_2) }\cdot \xi_{(\bb_2)} = 
\bb_1^2 [ 2  (1+ r^4)  +   2 r^2 \cos 2\p] , $

\noindent
so that these   vectors have a  finite norm for
a  flat disk ($e^{2\r}=1$) or  half-sphere ($e^{2\r}=4(1+ r^2)^{-2}$)
but their norm  formally diverges for $H^2$   ($e^{2\r}=4(1- r^2)^{-2}$).
One option then  is to regularize  the norms in  the same way  as  we do  for the  $H^2$   volume -- introduce a cutoff and drop power divergences.  We find 
(with a cutoff  at $r= e^{-\eps}$)  that   for the  $H^2$ volume 
 $\int^{ e^{-\eps}}_0 dr  {4 r \ov (1-r^2)^2} =  {1\ov \eps} - 1 + ... $    while   for the norms  
$\int^{ e^{-\eps}}_0 dr  {4^2 r^3 \ov (1-r^2)^4} =  {4\ov3 \eps^3} - {8\ov 3  \eps}   + {8\ov 3}  + ... $,  
$\int^{ e^{-\eps}}_0 dr  {4^2  2 r (1+r^4)  \ov (1-r^2)^4} =  {16\ov3 \eps^3} + {64\ov 3  \eps}   -  {64\ov 3}  + ... $.

\iffa 
{\tiny 
\begin{verbatim}
Remarks:
1.  what about growth of the $b_1,b_2$  modes  inside ads2 ($r\to 0$) ?  usually not allowed in D-case... 
the hope may be to  rule out these two  modes  and leave only $\rR$-one -- constant shifts along the circle. 
2. Just for the record and for future reference: if we just add to the Nambu action a term proportional to the boundary length, with s=ep cutoff in the (dt^2+ds^2)/sinh(s)^2 metric, we get
S_{bdy} = \int dt \sqrt{G_{mn} \dot X^m \dot X^n}|_{s=ep}
and after letting X0=t+xi0, X1 = s+xi1 and expanding to quadratic order in the longitudinal xi fluctuations:
S_{bdy} =\int dt [1/sinh(ep) - coth(ep)/sinh(ep) xi1
+1/(2sinh(ep)) (-xi1^2+2 Coth[s]^2 xi1^2-2 Coth[s] xi1 d_t xi0+(d_t xi1)^2) ]+...
The first term is the one that cancels the divergence in the classical area. But it looks like that if we plug in for the fluctuations the explicti expressions for the CKV's  (evaluated at s=ep)
xi^1 = sinh(s) (b1 cos(t)+b2 sin(t))
xi^0 = a  + cosh(s) (-b2 cos(t)+ b1 sin(t))  
then after integrating over t, it seems that everything cancels (it's not completely trivial and depends on the structure of the CKV), so these are still zero modes. 
So this doesn't seem to help to lift 2 of the CKV. 
(It's probably expected since the length should be invariant under the reparametrizations of the circle?). 
There is still the issue of normalizability, but it seems that no matter what we do with that, we will either keep all 3 or drop all 3 based on normalizability issues.  
\end{verbatim}
}
\fi 

As a side remark, let us comment on the possibility of having zero modes for the transverse $m^2=2$ fluctuation operator 
 in the AdS directions in  \rf{bb1},\rf{3}. If we focus on just a single transverse fluctuation within AdS$_3$, then one can formally find 3 zero modes related to the fact that the string solution breaks the $SO(3,1)$ isometries of AdS$_3$ down to $SO(2,1)$. Explicitly, taking the Poincar\'e coordinates on AdS$_3$ with metric $ds^2=\frac{1}{z^2}(dz^2+dr^2+r^2d\phi^2)$, the general AdS$_2$ string solution  ending on a circle    (of radius $\a$ and center at $(\b_1,\b_2)$)  at the boundary 
 can be written as 
\begin{equation}\la{bb8}
z^2+(r\cos\phi-\b_1)^2+(r\cos\phi-\b_2)^2=\a^2\,\ . 
\end{equation}
The parameters $\b_1$, $\b_2$ and $\a$  
correspond to broken translations and dilatation. The zero modes of the transverse fluctuation operator can be obtained as usual by taking derivatives of the classical solution with respect to these parameters. Expressing the result in the coordinates where the induced worldsheet metric is $ds^2=\frac{1}{\sinh^2\sigma}(d\sigma^2+d\tau^2)$ ($0<\tau<2\pi$, $\sigma>0$), the 3 zero modes are found to be
\begin{equation}\la{bb9}
\psi_{(\a)} = \coth\sigma\,,\qquad 
\psi_{{(\b_1)}}=\frac{\cos\tau}{\sinh\sigma}\,,\qquad 
\psi_{(\b_2)}=\frac{\sin\tau}{\sinh\sigma}\,.
\end{equation}
One can verify that these indeed satisfy 
\begin{equation}\la{bb10}
 \Big(\frac{\partial^2}{\partial \sigma^2}+\frac{\partial^2}{\partial\tau^2}
-\frac{2}{\sinh^2\sigma}\Big)\psi_{(\a,\b_1,\b_2)}=0\,.
\end{equation}
However, these zero modes are not normalizable. Moreover, they do not satisfy the 
Dirichlet boundary conditions at $\sigma=0$, as required for the transverse fluctuations, so they should not be relevant for our problem. Note also that, when considering all of the $n-2$ transverse directions in AdS$_n$, there would be, in fact,  $3(n-2)$ such zero modes (i.e., 9 in the AdS$_5$ case).


\end{document}

\bibitem{Marino:2016new} 
  M.~Marino,
  ``Localization at large N in Chern–Simons-matter theories,''
  J.\ Phys.\ A {\bf 50}, no. 44, 443007 (2017)
  [arXiv:1608.02959].